\documentclass[fleqn,usenatbib]{mnras}

\usepackage{newtxtext,newtxmath}

\usepackage[T1]{fontenc}

\DeclareRobustCommand{\VAN}[3]{#2}
\let\VANthebibliography\thebibliography
\def\thebibliography{\DeclareRobustCommand{\VAN}[3]{##3}\VANthebibliography}

\usepackage{graphicx}	
\usepackage{amsmath}	
\usepackage{orcidlink}

\usepackage{amsmath}
\usepackage{graphicx}
\usepackage{natbib}
\usepackage[scaled]{helvet}
\usepackage{epsfig}
\usepackage{url}
\bibpunct{(}{)}{;}{a}{}{,}
\newcommand{\kms}{\,km\,s$^{-1}$}
\usepackage{textcomp}
\usepackage{mathpazo}
\usepackage{xspace}
\usepackage{hyperref}
\usepackage{mathrsfs}
\usepackage{verbatim}

\usepackage{color}
\usepackage[normalem]{ulem}

\newcommand{\bjdtdb}{\ensuremath{\rm {BJD_{TDB}}}}
\newcommand{\feh}{\ensuremath{\left[{\rm Fe}/{\rm H}\right]}}

\newcommand{\teff}{\ensuremath{T_{\rm eff}}\xspace}

\newcommand{\msun}{\ensuremath{\,M_\odot}}
\newcommand{\rsun}{\ensuremath{\,R_\odot}}
\newcommand{\lsun}{\ensuremath{\,L_\odot}}
\newcommand{\mj}{\ensuremath{\,M_{\rm J}}}
\newcommand{\rj}{\ensuremath{\,R_{\rm J}}}

\newcommand{\Kepler}{{\it Kepler}}
\newcommand{\Ktwo}{{\it K2}}

\newcommand{\vsini}{\ensuremath{v\sin{I_*}}}
\newcommand{\be}{\begin{equation}}
\newcommand{\ee}{\end{equation}}

\newcommand{\tess}{{\it TESS}}
\newcommand{\TESS}{{\it TESS}}

\usepackage{lineno}
\newcommand{\cfa}{Center for Astrophysics \textbar \ Harvard \& Smithsonian, 60 Garden St, Cambridge, MA 02138, USA}
\newcommand{\msu}{Center for Data Intensive and Time Domain Astronomy, Department of Physics and Astronomy, Michigan State University, East Lansing, MI 48824, USA}

\newcommand{\utaustin}{Department of Astronomy, The University of Texas at Austin, Austin, TX 78712, USA}
\newcommand{\MIT}{Department of Physics and Kavli Institute for Astrophysics and Space Research, Massachusetts Institute of Technology, Cambridge, MA 02139, USA}
\newcommand{\MITEPS}{Department of Earth, Atmospheric and Planetary Sciences, Massachusetts Institute of Technology,  Cambridge,  MA 02139, USA}
\newcommand{\DOF}{Data Observatory Foundation, Chile}	
\newcommand{\Uminn}{University of Minnesota Duluth, Duluth, MN 55812, USA}

\newcommand{\riverside}{Department of Earth and Planetary Sciences, University of California, Riverside, CA 92521, USA}
\newcommand{\usq}{Centre for Astrophysics, University of Southern Queensland, West Street, Toowoomba, QLD 4350, Australia}
\newcommand{\ames}{NASA Ames Research Center, Moffett Field, CA, 94035, USA}

\newcommand{\warwick}{Department of Physics, University of Warwick, Gibbet Hill Road, Coventry CV4 7AL, UK}
\newcommand{\warwickceh}{Centre for Exoplanets and Habitability, University of Warwick, Gibbet Hill Road, Coventry CV4 7AL, UK}
\newcommand{\princeton}{Department of Astrophysical Sciences, Princeton University, 4 Ivy Lane, Princeton, NJ, 08544, USA}

\newcommand{\vanderbilt}{Department of Physics and Astronomy, Vanderbilt University, Nashville, TN 37235, USA}
\newcommand{\fisk}{Department of Physics, Fisk University, 1000 17th Avenue North, Nashville, TN 37208, USA}

\newcommand{\unc}{Department of Physics and Astronomy, University of North Carolina at Chapel Hill, Chapel Hill, NC 27599, USA}
\newcommand{\iac}{Instituto de Astrof\'isica de Canarias (IAC), E-38205 La Laguna, Tenerife, Spain}
\newcommand{\lalaguna}{Departamento de Astrof\'isica, Universidad de La Laguna (ULL), E-38206 La Laguna, Tenerife, Spain}
\newcommand{\louisville}{Department of Physics and Astronomy, University of Louisville, Louisville, KY 40292, USA}
\newcommand{\aavso}{American Association of Variable Star Observers 185 Alewife Brook Parkway, Suite 410 Cambridge, MA 02138}

\newcommand{\astrobiojapan}{Astrobiology Center, 2-21-1 Osawa, Mitaka, Tokyo 181-8588, Japan}
\newcommand{\ctio}{Cerro Tololo Inter-American Observatory, Casilla 603, La Serena, Chile}

\newcommand{\nexsci}{Caltech IPAC -- NASA Exoplanet Science Institute 1200 E. California Ave, Pasadena, CA 91125, USA}
\newcommand{\ucsc}{Department of Astronomy and Astrophysics, University of
California, Santa Cruz, CA 95064, USA}

\newcommand{\Pontificia}{Facultad de Ingeniería y Ciencias, Universidad Adolfo Ib\'a\~nez, Av. Diagonal las Torres 2640, Pe\~nalol\'en, Santiago, Chile}
\newcommand{\Millennium}{Millennium Institute for Astrophysics, Chile}
\newcommand{\maxplank}{Max-Planck-Institut f\"ur Astronomie, K\"onigstuhl 17, Heidelberg 69117, Germany}

\newcommand{\osu}{Department of Astronomy, The Ohio State University, 140 West 18th Avenue, Columbus, OH 43210, USA}
\newcommand{\MITAA}{Department of Aeronautics and Astronautics, MIT, 77 Massachusetts Avenue, Cambridge, MA 02139, USA}

\newcommand{\lehigh}{Department of Physics, Lehigh University, 16 Memorial Drive East, Bethlehem, PA 18015, USA}
\newcommand{\utah}{Department of Physics and Astronomy, University of Utah, 115 South 1400 East, Salt Lake City, UT 84112, USA}

\newcommand{\UPenn}{The University of Pennsylvania, Department of Physics and Astronomy, Philadelphia, PA, 19104, USA}
\newcommand{\montana}{Department of Physics and Astronomy, University of Montana, 32 Campus Drive, No. 1080, Missoula, MT 59812 USA}
\newcommand{\psu}{Department of Astronomy \& Astrophysics, The Pennsylvania State University, 525 Davey Lab, University Park, PA 16802, USA}
\newcommand{\psust}{Center for Exoplanets and Habitable Worlds, The Pennsylvania State University, 525 Davey Lab, University Park, PA 16802, USA}
\newcommand{\PSUET}{Penn State Extraterrestrial Intelligence Center, 525 Davey Laboratory, The Pennsylvania State University, University Park, PA, 16802, USA}
\newcommand{\Kutztown}{Department of Physical Sciences, Kutztown University, Kutztown, PA 19530, USA}

\newcommand{\saao}{South African Astronomical Observatory, PO Box 9, Observatory, 7935, Cape Town, South Africa}
\newcommand{\salt}{Southern African Large Telescope, PO Box 9, Observatory, 7935, Cape Town, South Africa}

\newcommand{\txamGP}{George P.\ and Cynthia Woods Mitchell Institute for Fundamental Physics and Astronomy, Texas A\&M University, College Station, TX77843 USA}

\newcommand{\wellesley}{Department of Astronomy, Wellesley College, Wellesley, MA 02481, USA}

\newcommand{\byu}{Department of Physics and Astronomy, Brigham Young University, Provo, UT 84602, USA}

\newcommand{\eso}{European Southern Observatory, Alonso de C\'ordova 3107, Vitacura, Casilla 19001, Santiago, Chile}

\newcommand{\keele}{Astrophysics Group, Keele University, Staffordshire ST5 5BG, UK}

\newcommand{\gmu}{George Mason University, 4400 University Drive MS 3F3, Fairfax, VA 22030, USA}
\newcommand{\unsw}{Exoplanetary Science at UNSW, School of Physics, UNSW Sydney, NSW 2052, Australia}
\newcommand{\sifa}{School of Physics, Sydney Institute for Astronomy (SIfA), The University of Sydney, NSW 2006, Australia}
\newcommand{\nanjing}{School of Astronomy and Space Science, Key Laboratory of Modern Astronomy and Astrophysics in Ministry of Education, Nanjing University, Nanjing 210046, Jiangsu, China}
\newcommand{\berkely}{Department of Astronomy, University of California Berkeley, Berkeley, CA 94720-3411, USA}

\newcommand{\Patashnick}{Patashnick Voorheesville Observatory, Voorheesville, NY 12186, USA}

\newcommand{\Tsinghua}{Department of Astronomy, Tsinghua University, Beijing 100084, China}

\newcommand{\wisconsin}{Department of Astronomy, University of Wisconsin-Madison, Madison, WI 53706, USA}

\newcommand{\ASTRAVEO}{ASTRAVEO, LLC, PO Box 1668, Gloucester, MA 01931}
\newcommand{\TJHS}{Thomas Jefferson High School, 6560 Braddock Rd, Alexandria, VA 22312 USA}

\newcommand{\gemini}{Gemini Observatory/NSF’s NOIRLab, 670 N. A’ohoku Place, Hilo, HI, 96720, USA}
\newcommand{\umd}{Department of Astronomy, University of Maryland, College Park, College Park, MD}
\newcommand{\ucscchile}{Departamento de Matem\'atica y F\'isica Aplicadas, Facultad de Ingenier\'ia, Universidad Cat\'olica de la Sant\'isima Concepci\'on, Alonso de Rivera 2850, Concepci\'on, Chile }
\newcommand{\Kotizarovci}{Kotizarovci Observatory, Sarsoni 90, 51216 Viskovo, Croatia}
\newcommand{\Villa}{Villa ’39 Observatory, Landers, CA 92285, USA}
\newcommand{\ElSauce}{El Sauce Observatory, Coquimbo Province, Chile}

\newcommand{\calou}{Observatri de Ca l'Ou}
\newcommand{\Sternberg}{Sternberg Astronomical Institute, M.V. Lomonosov Moscow State University, 13, Universitetskij pr., 119234, Moscow, Russia}
\newcommand{\Ural}{Ural Federal University, Ekaterinburg, Russia, ul. Mira d. 19, Yekaterinburg, Russia, 620002}
\newcommand{\Oikaimeden}{Oukaimeden Observatory, High Energy Physics and Astrophysics Laboratory, Cadi Ayyad University, Marrakech, Morocco}
\newcommand{\liegeastrobio}{Astrobiology Research Unit, Universit\'e de Li\`ege, 19C All\'ee du 6 Ao\^ut, 4000 Li\`ege, Belgium}
\newcommand{\Faculdade}{Departamento de f\'isica, e Astronomia, Faculdade de Ci\`encias, Universidade do Porto, Rua Campo Alegre, 4169-007, Porto, Portugal}
\newcommand{\CAUP}{Instituto de Astrof\'isica e Ci\`encias do Espa\c{c}o, CAUP, Universidade do Porto, Rua das Estrelas, 4150-762, Porto, Portugal}
\newcommand{\KASSI}{Korea Astronomy and Space Science Institute, 776 Daedeok-daero, Yuseong-gu, Daejeon 34055, Republic of Korea}
\newcommand{\Shanghai}{Shanghai Astronomical Observatory, Chinese Academy of Sciences, Shanghai 200030, China}
\newcommand{\SFASU}{Department of Physics, Engineering and Astronomy, Stephen F. Austin State University, TX 75962, USA}
\newcommand{\albany}{Observatori Astron\^{o}mic Albany\`{a}, Cam\'i de Bassegoda S/N, Albany\`{a} 17733, Girona, Spain}
\newcommand{\swinburne}{Swinburne University of Technology, Centre for Astrophysics and Supercomputing, John Street, Hawthorn, VIC 3122, Australia}
\newcommand{\bern}{Physikalisches Institut, University of Bern, Gesellschaftsstrasse 6, 3012 Bern, Switzerland}
\newcommand{\MunnerlynLab}{Charles R. \& Judith G. Munnerlyn Astronomical Laboratory, Department of Physics \& Astronomy, Texas A\&M University, College Station, TX 77843, USA}
\newcommand{\CGWA}{Center for Gravitational Wave Astronomy, The University of Texas Rio Grande Valley, Brownsville, TX 78520, USA}
\newcommand{\Komaba}{Komaba Institute for Science, The University of Tokyo, 3-8-1 Komaba, Meguro, Tokyo 153-8902, Japan}
\newcommand{\waffelow}{Waffelow Creek Observatory}
\newcommand{\Cadi}{Fundamental and Applied Physics Laboratory, Physics Department, Polydisciplinary Faculty of Safi, Cadi Ayyad University, Marrakesh, Morocco}
\newcommand{\maurylewin}{The Maury Lewin Astronomical Observatory, Glendora,California.91741. USA}
\newcommand{\citizen}{Citizen Scientist}

\title[Cargo Ship II: 6 Planets Delivery]{Another Shipment of Six Short-Period Giant Planets from {\it TESS}}
\vspace{-0.5in}

\author[J. E. Rodriguez et al.]{Joseph E. Rodriguez,$^{1\thanks{E-mail: jrod@msu.edu }\orcidlink{0000-0001-8812-0565}}$
Samuel N. Quinn,$^{2\orcidlink{0000-0002-8964-8377}}$
Andrew Vanderburg,$^{3\orcidlink{0000-0001-7246-5438}}$
George Zhou,$^{4\orcidlink{0000-0002-4891-3517}}$
Jason D. Eastman,$^{2\orcidlink{0000-0003-3773-5142}}$
\newauthor
Erica Thygesen,$^{1\orcidlink{0000-0002-9165-6245}}$
Bryson Cale,$^{5\orcidlink{0000-0001-6279-0595}}$
David R. Ciardi,$^{5\orcidlink{0000-0002-5741-3047}}$
Phillip A.\ Reed,$^{6\orcidlink{0000-0002-5005-1215}}$ 
Ryan J. Oelkers,$^{7,8,9\orcidlink{0000-0002-0582-1751}}$ 
\newauthor
Karen A.\ Collins,$^{2\orcidlink{0000-0001-6588-9574}}$
Allyson Bieryla,$^{2\orcidlink{0000-0001-6637-5401}}$
David W. Latham,$^{2\orcidlink{0000-0001-9911-7388}}$
Erica J.\ Gonzales,$^{10\orcidlink{0000-0002-9329-2190}}$
B. Scott Gaudi,$^{11\orcidlink{0000-0003-0395-9869}}$
\newauthor
Coel Hellier,$^{12}$
Mat\'ias I. Jones$^{13\orcidlink{0000-0002-9158-7315}}$
Rafael Brahm$^{14, 15, 16\orcidlink{0000-0002-9158-7315}}$
Kirill Sokolovsky,$^{1\orcidlink{0000-0001-5991-6863}}$ 
Jack Schulte,$^{1\orcidlink{0000-0002-7382-0160}}$ 
Gregor Srdoc,$^{17}$ 
\newauthor
John Kielkopf,$^{18\orcidlink{0000-0003-0497-2651}}$
Ferran Grau Horta,$^{19\orcidlink{0000-0001-9927-7269}}$
Bob Massey,$^{20\orcidlink{0000-0001-8879-7138}}$
Phil Evans,$^{21\orcidlink{0000-0002-5674-2404}}$
Denise C. Stephens,$^{22\orcidlink{0000-0003-4658-7567}}$
\newauthor
Kim K.\ McLeod,$^{23\orcidlink{0000-0001-9504-1486}}$
Nikita Chazov,$^{24\orcidlink{0000-0002-4070-7831}}$
Vadim Krushinsky,$^{25\orcidlink{0000-0001-9388-691X}}$
Mourad Ghachoui,$^{26,27\orcidlink{0000-0001-9388-691X}}$
Boris S. Safonov,$^{28\orcidlink{0000-0003-1713-3208}}$
\newauthor
Cayla M. Dedrick,$^{29,30\orcidlink{0000-0001-9408-8848}}$
Dennis Conti,$^{30\orcidlink{0000-0003-2239-0567}}$ 
Didier Laloum,$^{30}$
Steven Giacalone,$^{31\orcidlink{0000-0002-8965-3969}}$
Carl Ziegler,$^{32\orcidlink{0000-0002-0619-7639}}$
\newauthor
Pere Guerra Serra,$^{33\orcidlink{0000-0002-4308-2339}}$
Ramon Naves Nogues,$^{34\orcidlink{0000-0002-9463-9029}}$
Felipe Murgas,$^{35,36\orcidlink{0000-0001-9087-1245}}$
Edward J. Michaels,$^{37\orcidlink{0000-0001-8746-4358}}$
\newauthor
George R. Ricker,$^{3\orcidlink{0000-0003-2058-6662}}$
Roland K. Vanderspek,$^{3}$
Sara Seager,$^{3,38,39}$
Joshua N. Winn,$^{40\orcidlink{0000-0002-6892-6948}}$
Jon M. Jenkins,$^{41\orcidlink{0000-0002-4715-9460}}$
\newauthor
Brett Addison,$^{42, 4\orcidlink{0000-0003-3216-0626}}$
Owen Alfaro,$^{43}$
D. R. Anderson,$^{13, 44}$
Elias Aydi,$^{1\orcidlink{0000-0001-8525-3442}}$ 
Thomas G.\ Beatty,$^{45,\orcidlink{0000-0002-9539-4203}}$
\newauthor
Timothy R. Bedding,$^{46,\orcidlink{0000-0001-5222-4661}}$ 
Alexander A.\ Belinski,$^{27\orcidlink{0000-0003-3469-0989}}$ 
Zouhair Benkhaldoun,$^{25\orcidlink{0000-0001-6285-9847}}$ 
Perry Berlind,$^{2}$ 
\newauthor
Cullen H. Blake,$^{47\orcidlink{0000-0002-6096-1749}}$%
Michael J. Bowen,$^{43}$
Brendan P. Bowler,$^{48}$  
Andrew W. Boyle,$^{5\orcidlink{0000-0001-6037-2971}}$%
Dalton Branson,$^{49}$
\newauthor
C\'{e}sar Brice\~{n}o,$^{50}$
Michael L. Calkins,$^{2\orcidlink{0000-0002-2830-5661}}$
Emma Campbell,$^{23}$
Jessie L. Christiansen,$^{5\orcidlink{0000-0002-8035-4778}}$
Laura Chomiuk,$^{1, \orcidlink{0000-0002-8400-3705}}$
\newauthor
Kevin I. Collins,$^{43\orcidlink{0000-0003-2781-3207}}$
Matthew A. Cornachione,$^{51\orcidlink{0000-0003-1012-4771}}$%
Ahmed Daassou,$^{52\orcidlink{0000-0001-9439-5047}}$
Courtney D. Dressing,$^{31\orcidlink{0000-0001-8189-0233}}$%
\newauthor
Gilbert A. Esquerdo,$^{2\orcidlink{0000-0002-9789-5474}}$%
Dax L. Feliz,$^{53\orcidlink{0000-0002-2457-7889}}$%
William Fong,$^3$
Akihiko Fukui,$^{54,35\orcidlink{0000-0002-4909-5763}}$
Tianjun Gan,$^{55\orcidlink{0000-0002-4503-9705}}$%
Holden Gill,$^{31\orcidlink{0000-0001-6171-7951}}$%
\newauthor
Maria V. Goliguzova,$^{27\orcidlink{0000-0003-2228-7914}}$
Jarrod Hansen,$^{22\orcidlink{0000-0002-1321-3174}}$%
Thomas Henning,$^{56}$
Eric G. Hintz,$^{19\orcidlink{0000-0002-9867-7938}}$%
Melissa J.\ Hobson,$^{56, 15\orcidlink{0000-0002-5945-7975}}$%
\newauthor
Jonathan Horner,$^{4\orcidlink{0000-0002-1160-7970}}$
Chelsea X. Huang,$^{4\orcidlink{0000-0003-0918-7484}}$
David J. James,$^{57\orcidlink{0000-0001-5160-4486}}$
Jacob S. Jensen,$^{21}$
Samson A. Johnson,$^{11\orcidlink{0000-0001-9397-4768}}$
\newauthor
Andr\'es Jord\'an,$^{14,15\orcidlink{0000-0002-9158-7315}}$
Stephen R. Kane,$^{58}$
Khalid Barkaoui,$^{26,38,35\orcidlink{0000-0003-1464-9276}}$
Myung-Jin Kim,$^{59\orcidlink{0000-0002-4787-6769}}$
Kingsley Kim,$^{60}$
\newauthor
Rudolf B. Kuhn,$^{61,62\orcidlink{0000-0002-4236-9020}}$
Nicholas Law,$^{63\orcidlink{0000-0001-9380-6457}}$
Pablo Lewin,$^{64\orcidlink{0000-0003-0828-6368}}$
Hui-Gen Liu,$^{65\orcidlink{0000-0001-5162-1753}}$
Michael B. Lund,$^{5\orcidlink{0000-0002-4236-9020}}$
\newauthor
Andrew W. Mann,$^{63\orcidlink{0000-0003-3654-1602}}$
Nate McCrady,$^{49\orcidlink{0000-0002-8041-1832}}$
Matthew W. Mengel,$^{4\orcidlink{0000-0002-7830-6822}}$ 
Jessica Mink,$^{2\orcidlink{0000-0003-3594-1823}}$
Lauren G. Murphy,$^{6\orcidlink{0000-0003-3796-6303}}$
\newauthor
Norio Narita,$^{35,65,66\orcidlink{0000-0001-8511-2981}}$
Patrick Newman,$^{42\orcidlink{0000-0003-3848-3418}}$
Jack Okumura,$^{4\orcidlink{0000-0002-4876-8540}}$ 
Hugh~P.~Osborn,$^{3,67\orcidlink{0000-0002-4047-4724}}$
Martin Paegert,$^{2\orcidlink{0000-0001-8120-7457}}$
\newauthor
Enric Palle,$^{35,36\orcidlink{0000-0003-0987-1593}}$
Joshua Pepper,$^{68\orcidlink{0000-0002-3827-8417}}$
Peter Plavchan,$^{43\orcidlink{0000-0002-8864-1667}}$
Alexander A. Popov,$^{24\orcidlink{0000-0002-2007-2461}}$
Markus Rabus,$^{69\orcidlink{0000-0003-2935-7196}}$
\newauthor
Jessica Ranshaw,$^{1}$
Jennifer A. Rodriguez,$^{1\orcidlink{0000-0003-1560-001X}}$
Dong-Goo Roh,$^{65\orcidlink{0000-0001-6104-4304}}$
Michael A. Reefe,$^{3\orcidlink{0000-0003-4701-8497}}$
Arjun B. Savel,$^{70\orcidlink{0000-0002-2454-768X}}$
\newauthor
Richard P. Schwarz,$^{71\orcidlink{0000-0001-8227-1020}}$
Avi Shporer,$^{3\orcidlink{0000-0002-1836-3120}}$
Robert J.\ Siverd,$^{72\orcidlink{0000-0001-5016-3359}}$
David H. Sliski,$^{47}$
Keivan G. Stassun,$^{53,73\orcidlink{0000-0002-3481-9052}}$%
\newauthor
Daniel J.\ Stevens,$^{74\orcidlink{0000-0002-5951-8328}}$%
Abderahmane Soubkiou,$^{23,75,76}$
Eric B. Ting,$^{41}$
C.G. Tinney,$^{77}$
Noah Vowell,$^{1}$
\newauthor
Payton Walton,$^{1}$
R. G. West,$^{44,78}$
Maurice L. Wilson,$^{2\orcidlink{0000-0003-1928-0578}}$%
Robert A. Wittenmyer,$^{4\orcidlink{0000-0001-9957-9304}}$%
Justin M. Wittrock,$^{43\orcidlink{0000-0002-7424-9891}}$
\newauthor
Shania Wolf,$^{67}$
Jason T. Wright,$^{28,29,79\orcidlink{0000-0001-6160-5888}}$%
Hui Zhang,$^{80\orcidlink{0000-0003-3491-6394}}$
and Evan Zobel$^{1}$
\\
\emph{\normalsize Affiliations are listed at the end of the paper}
}

\date{Accepted XXX. Received YYY; in original form ZZZ}

\pubyear{2022}

\begin{document}

\label{firstpage}
\pagerange{\pageref{firstpage}--\pageref{lastpage}}
\maketitle


\clearpage
\begin{abstract}
We present the discovery and characterization of six short-period, transiting giant planets from NASA’s {\it Transiting Exoplanet Survey Satellite ({\it TESS})} --- TOI-1811 (TIC 376524552), TOI-2025 (TIC 394050135), TOI-2145 (TIC 88992642), TOI-2152 (TIC 395393265), TOI-2154 (TIC 428787891), \& TOI-2497 (TIC 97568467). All six planets orbit bright host stars (8.9 $<G<$ 11.8, 7.7 $<K<$ 10.1). Using a combination of time-series photometric and spectroscopic follow-up observations from the {\it TESS} Follow-up Observing Program (TFOP) Working Group, we have determined that the planets are Jovian-sized (R$_{\rm P}$ = 0.99-1.45 R$_{\rm J}$), have masses ranging from 0.92 to 5.26 M$_{\rm J}$, and orbit F, G, and K stars (4766 $\le$ T$_{\rm eff}$ $\le$ 7360 K). We detect a significant orbital eccentricity for the three longest-period systems in our sample: TOI-2025 b (P = 8.872 days, 0.394$^{+0.035}_{-0.038}$), TOI-2145 b (P = 10.261 days, $e$ = $0.208^{+0.034}_{-0.047}$), and TOI-2497 b (P = 10.656 days, $e$ = $0.195^{+0.043}_{-0.040}$). TOI-2145 b and TOI-2497 b both orbit subgiant host stars (3.8 $<$ $\log$ g $<$4.0), but these planets show no sign of inflation despite very high levels of irradiation. The lack of inflation may be explained by the high mass of the planets; $5.26^{+0.38}_{-0.37}$ M$_{\rm J}$ (TOI-2145 b) and $4.82\pm0.41$ M$_{\rm J}$ (TOI-2497 b). These six new discoveries contribute to the larger community effort to use {\it TESS} to create a magnitude-complete, self-consistent sample of giant planets with well-determined parameters for future detailed studies. 

\end{abstract}

\begin{keywords}
techniques: radial velocities -- techniques: photometric -- planets and satellites: detection
\end{keywords}

\section{Introduction}
While NASA's {\it Transiting Exoplanet Survey Satellite} (\tess) mission continues to discover a wealth of new small planets, it is also discovering many transiting hot and warm Jupiters, complementing the prior work of ground-based transit surveys \citep{Pollacco:2006, Pepper:2007, Bakos:2013} and space-based surveys like NASA's \Kepler\ and \Ktwo\ missions \citep{Borucki:2010, Howell:2014} and ESA's CoRoT satellite \citep{Auvergne:2009}. These surveys discovered hundreds of hot Jupiters and established that they are rare ($<$1\%). Using observations from \Kepler, three different occurrence rates of hot Jupiters have been measured: 0.43$\pm$0.05\% \citep{Fressin:2013}, 0.57$^{+0.14}_{-0.12}$\% \citep{Petigura:2018}, and 0.43$^{+0.07}_{-0.06}$\% \citep{Masuda:2017}. However, radial velocity (RV) surveys have measured the occurrence rate to be significantly higher: 1.5$\pm$0.6\% \citep{Cumming:2008} and 1.2$\pm$0.4\% \citep{Wright:2012}, with the difference in occurrence rates possibly due to the removal of spectroscopic binaries (SB2 that show two sets of lines and short-period SB1s where only one set of lines is detected but with a large RV offset consistent with a stellar companion) in the RV surveys \citep{Moe:2021}.  Since the surveys have different target selection criteria, these results suggest that the occurrence rates depend on the properties of the host star (mass, multiplicity, age, etc). \citet{Zhou:2019} gave a first glimpse into the occurrence rate from the primary mission of NASA's \tess\ \citep{Ricker:2015}), measuring an occurrence rate of 0.41$\pm$0.10\%, consistent with results from the \Kepler\ mission. \citet{Zhou:2019} used \tess\ data to measure occurrence rates as a function of spectral type and found it to be $0.71\pm0.31\%$ for G stars, $0.43\pm0.15\%$ for F stars, and $0.26\pm0.11\%$ for A stars.

\begin{figure*}
\centering 
\includegraphics[trim = 0 0 0 0,width=\linewidth]{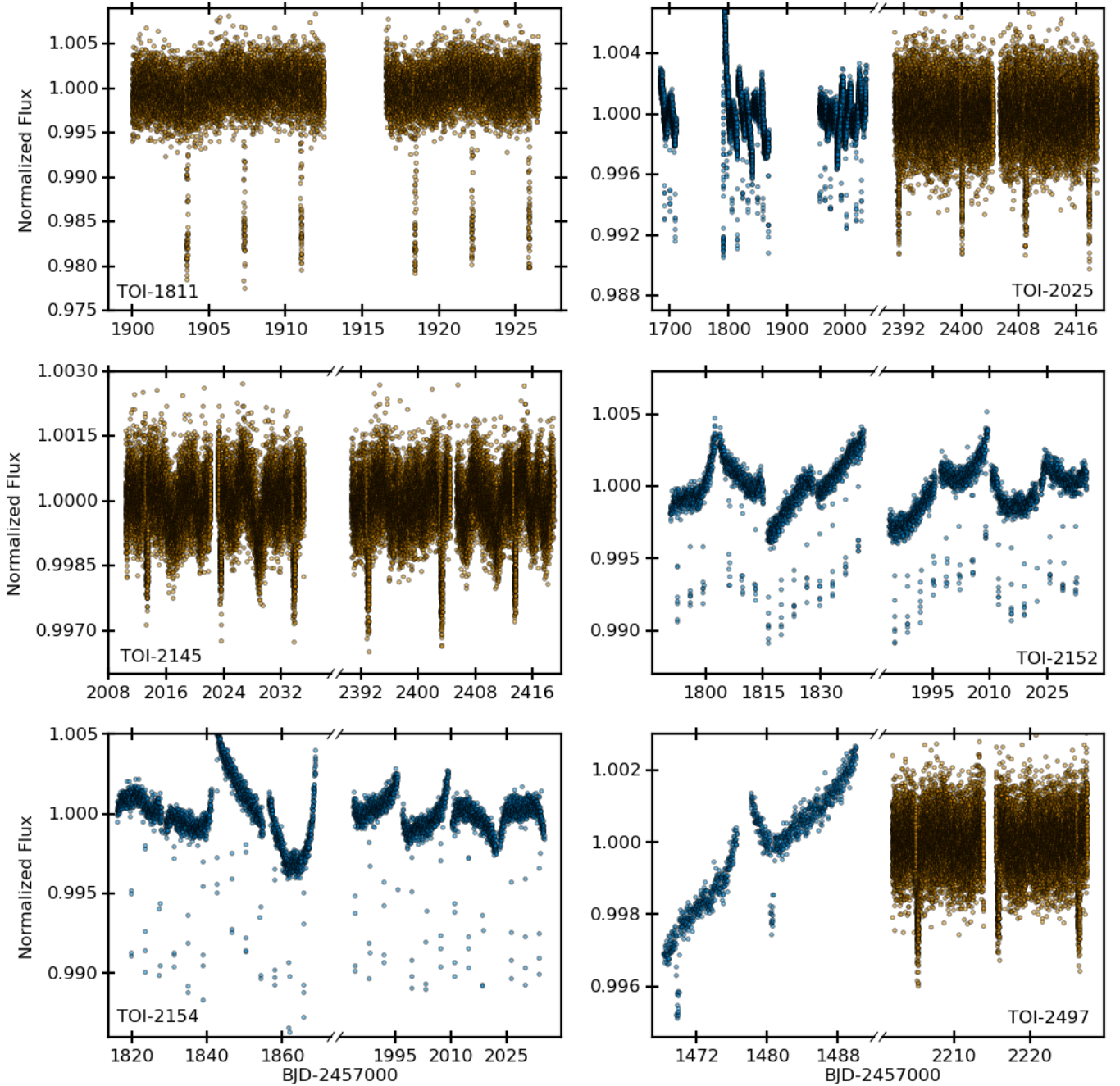}
\caption{The \tess\ 30-minute light curves extracted using the technique described in \S\ref{sec:TESS} (blue) and 2-minute SPOC light curves (orange) for TOI-1811 (top-left), TOI-2025 (top-right), TOI-2145 (middle-left), TOI-2152 (middle-right), TOI-2154 (bottom-left), and TOI-2497 (bottom-right). }
\label{fig:fullLCs}
\end{figure*}

As a result of its observing strategy and photometric precision, \tess\ should be nearly complete for discovering transiting hot Jupiters (P$<$10 days, \tess$_{\rm Mag}$ $<$ 10, \citealp{Zhou:2019}), providing the community with the opportunity to create a homogeneous, magnitude-complete population of giant planet parameters. Unfortunately, most ground-based surveys struggled to discover transiting planets with periods above $\sim$5 days due to their poor duty cycle \citep{Gaudi:2005}. However, much work remains as recent results suggest that the current sample of known hot Jupiters is only 75\% complete for stars brighter than Gaia magnitude \citep{Gaia:2018} G$\le$10.5, 50\% for G$\le$12, and 36\% at G$\le$12.5 \citep{Yee:2021}. Fortunately, coordinated RV efforts within the \TESS\ Follow-up Observing Program (TFOP) are helping to extend this sample to $G<$12.5. As we continue to confirm new hot Jupiters from \tess, we will gain insight into some of the key questions about their formation and evolutionary pathways \citep[see reviews, e.g.,][]{Dawson:2018, Fortney:2021}.

\begin{figure}
\centering 
\includegraphics[trim = 0 0 0 0,width=\linewidth]{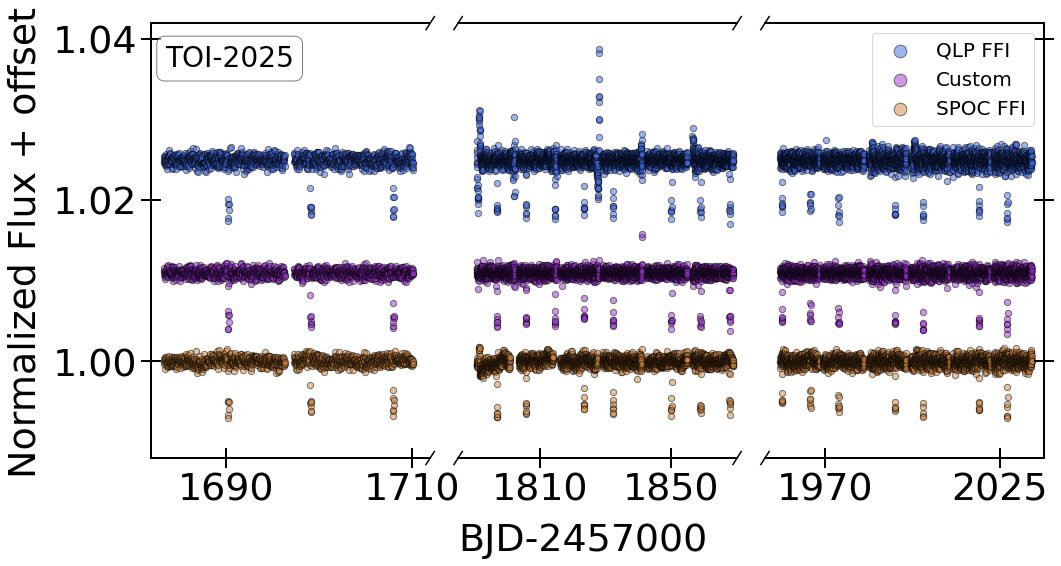}
\caption{The flattened \tess\ 30-minute light curves reduced using with the Quick Look Pipeline (Blue, \citealp{Huang:QLP}), Custom (purple, \citealp{Vanderburg:2019}), and flattened \tess-SPOC light curves generated from the FFIs (Orange, \citealp{Jenkins:2016, Caldwell:2020}) for TOI-2025. These lightcurves are flattened using {\it Keplerspline} (as described in \S\ref{sec:TESS}) prior to being included in the global fit (see \S\ref{sec:GlobalModel}).}
\label{fig:LCcomp}
\end{figure}

Here we present the discovery and characterization of six new hot and warm giant planets from NASA's \tess\ mission. These six targets were selected for follow up confirmation as part of a large effort to discover and characterize transiting hot and warm Jupiters with the goal of creating a magnitude-complete sample of giant planets with measured eccentricities \citep{Rodriguez:2019, Rodriguez:2021, IkwutUkwa:2022}. These discoveries, combined with other large scale efforts to use \tess\ to confirm and characterize giant planets (\citealp{Nielsen:2019, Brahm:2020, Addison:2021, Grunblatt:2022}, Yee et al. submitted), should lead to a magnitude-complete sample of hot Jupiters for future population studies. During the preparation of this paper, we became aware of another effort to announce the discovery of TOI-2025 b \citep{Knudstrup:2022}. Future efforts should combine all observations of TOI-2025 b presented in both discovery papers.  All results presented here on TOI-2025 were independently determined, and all communication between both groups was related to coordinating submissions. In \S2 we present the \tess\ and follow-up observations. We review our global analysis using \texttt{EXOFASTv2} \citep{Eastman:2019} in \S3 and discuss our results in \S4, specifically the impact \tess\ is having on our understanding of hot Jupiters. Our conclusions for this work are summarized in \S5.

\begin{figure*}
	\centering\vspace{.0in}
	\includegraphics[width=0.9\linewidth, trim={0 0 0 0}, clip]{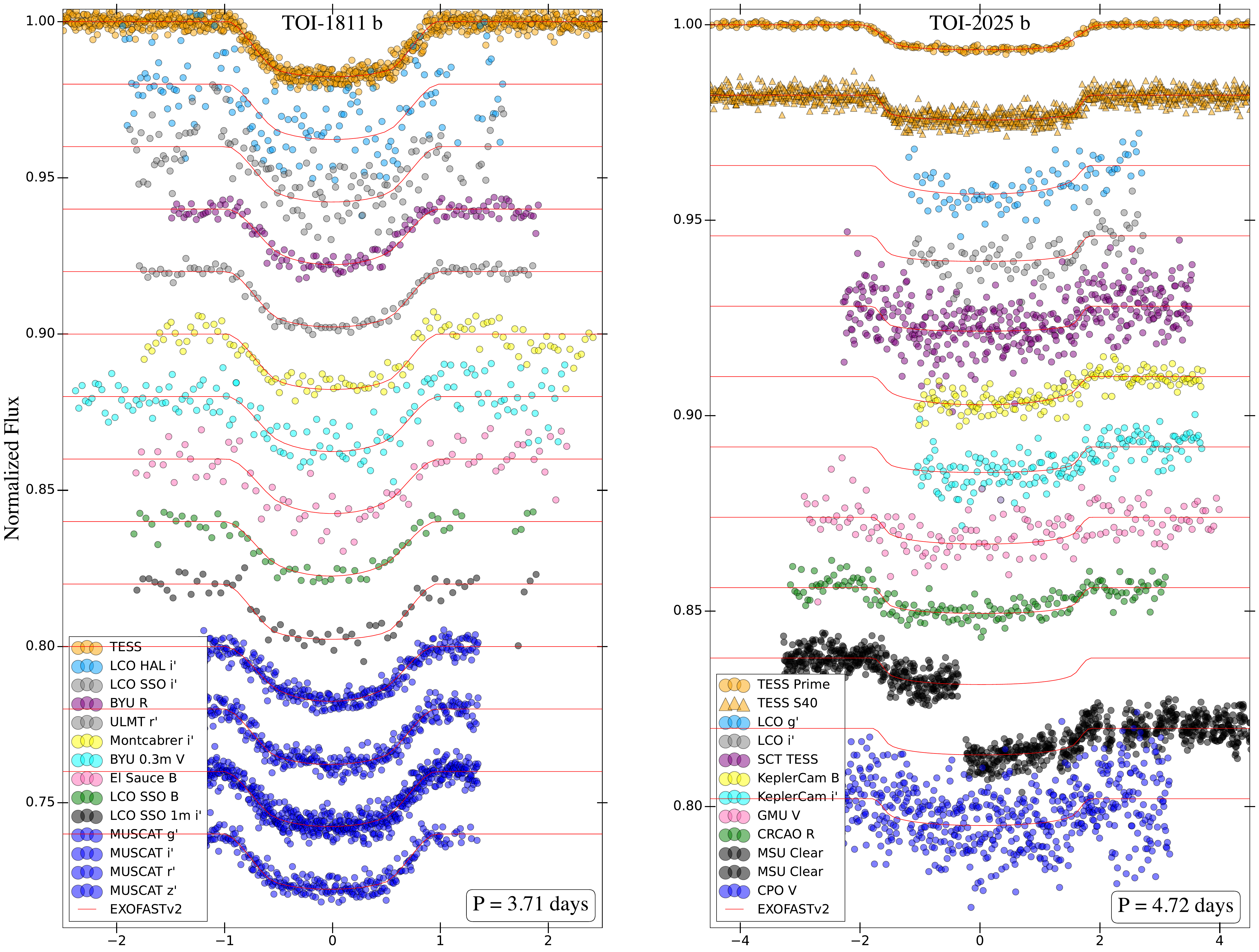}\\
	\includegraphics[width=0.9\linewidth, trim={0 0 0 0}, clip]{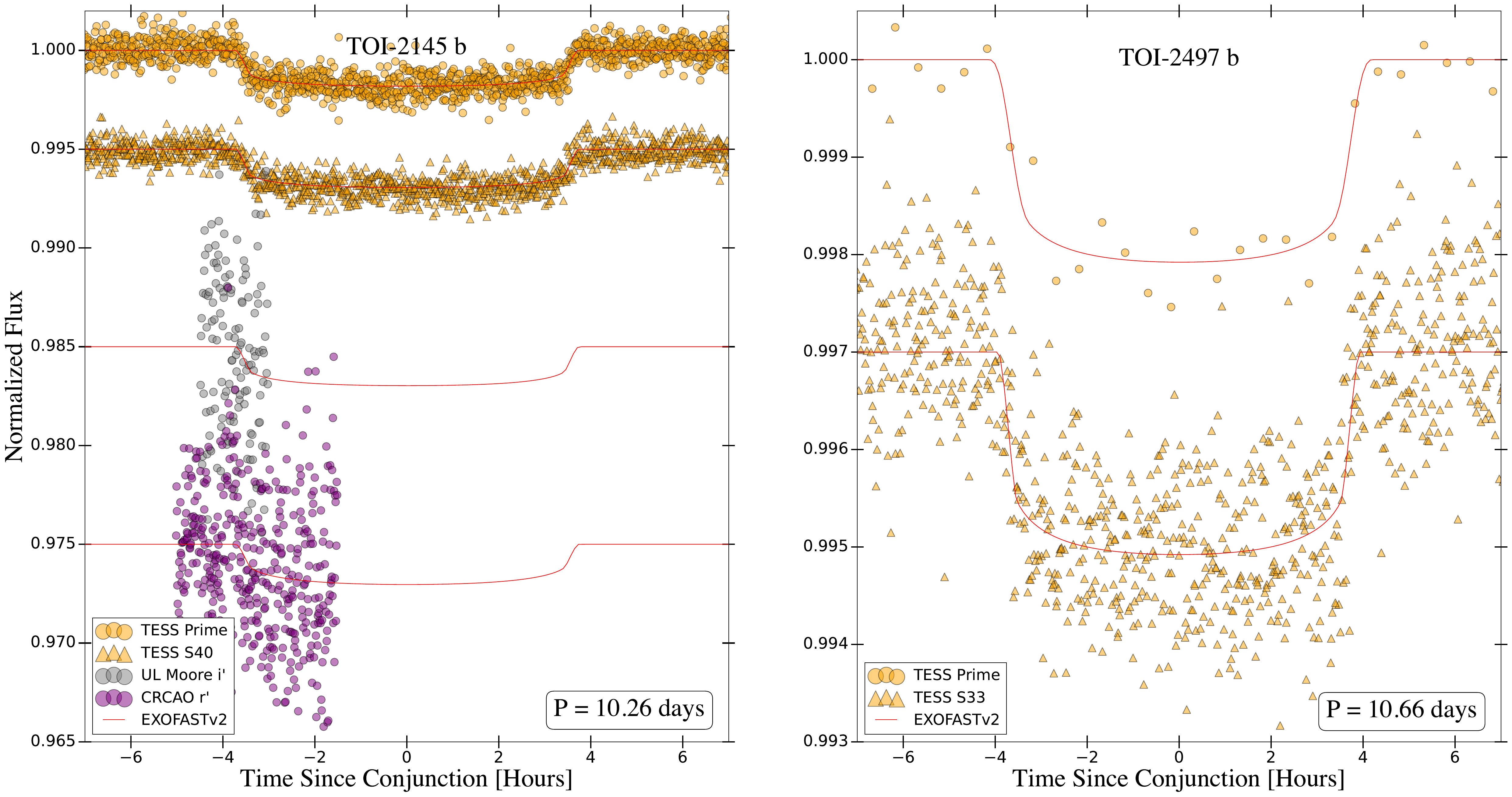}\\
	\caption{The \tess\ (orange) and TFOP SG1 follow-up transits of TOI-1811 b (top-left), TOI-2025 b (top-right), TOI-2145 b (bottom-left), and TOI-2497 b (bottom-right). The \texttt{EXOFASTv2} model for each transit observation is shown by the red solid line. }
	\label{fig:transits1} 
\end{figure*}

\begin{figure*}
	\centering\vspace{.0in}
	\includegraphics[width=0.9\linewidth, trim={0 0 0 0}, clip]{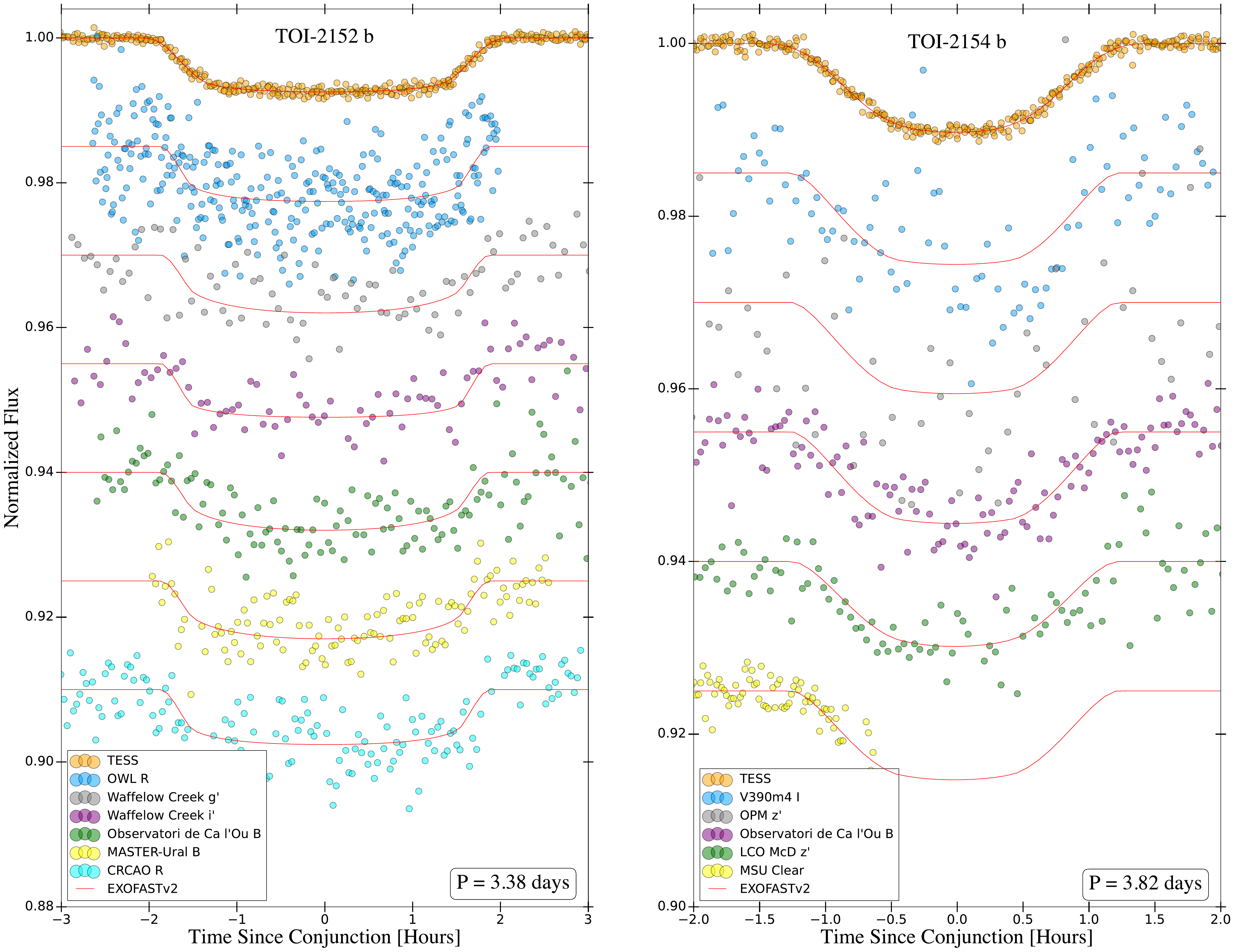}\\
	\caption{The \tess\ (orange) and TFOP SG1 follow-up transits of TOI-2152A b (Left) and TOI-2154 b (Right). The \texttt{EXOFASTv2} model for each transit observation is shown by the red solid line.}
	\label{fig:transits2} 
\end{figure*}

\vspace{-0.25in}
\section{Observations and Archival Data}
\label{sec:Obs}
We used a series of photometric and spectroscopic observations to rule out false positive scenarios, confirm planet candidates as bona fide planets, and measure key parameters such as orbital eccentricity and the planet's mass. All observations presented here were coordinated through the {\it TESS} Follow-up Observing Program (TFOP) Working Groups. The literature values for previously measured parameters of these stars are listed in Table \ref{tbl:LitProps}.

\subsection{{\it TESS} Photometry}
\label{sec:TESS}
Launched in 2018, NASA's \tess\ mission has been in full operation with over 200 planets confirmed to date\footnote{\url{https://exoplanetarchive.ipac.caltech.edu/}}. Using a 24$^\circ$x96$^\circ$ field of view, \TESS\ monitors each observing sector for $\sim$27 days before moving to the next sector \citep{Ricker:2015}. During the prime mission, \tess\ observed nearly the entire sky at a 30-minute cadence and a pre-selected set of a few hundred thousand stars at 2-minute cadence. After a successful 2-year primary mission that observed each ecliptic hemisphere for about a year, \tess\ began its 27-month first extended mission that is ongoing and has already revisited some of the prime-mission targets but also observed a large portion of the ecliptic plane, where the repurposed \Kepler\ mission (\Ktwo, \citealp{Howell:2014}) discovered over 500 planetary systems and over 1000 more candidates \citep[e.g.]{Barros:2016, Crossfield:2016, Vanderburg:2016b, Mayo:2018, Zink:2019, HardegreeUllman:2020, Zink:2021}. During the 27-month extended mission, \tess\ has added a third, 20-second cadence mode for some pre-selected targets and the exposure time of the Full Frame Images (FFI, where the entire 24$^\circ$x96$^\circ$ field of view is observed) was reduced to 10 minutes. To date, \TESS\ has announced over 5000 targets that display a signal consistent with it being an exoplanet, which are known as \TESS\ Objects of Interest\footnote{\url{https://tess.mit.edu/toi-releases/}} (TOIs, \citealp{Guerrero:2021}), targets that display a signal consistent with it being an exoplanet. 

\TESS\ observed all six TOIs presented here during the 2-year primary mission, and, in the cases of TOI-2025 and TOI-2497, reobserved during the extended mission. TOI-1811 and TOI-2145 were only observed at 2-minute cadence, TOI-2152 and TOI-2154 were only observed in the 30-min full frame images, and TOI-2025 and TOI-2497 were observed in both cadences during different sectors (see Figure \ref{fig:fullLCs}). For the 2-minute observations, the \tess\ images were downlinked, reduced, and analyzed by the Science Processing Operations Center (SPOC) pipeline \citep{Smith:2012, Stumpe:2014, Jenkins:2016}. The final SPOC lightcurves were searched for transits with the SPOC Transiting Planet Search (TPS, \citealt{Jenkins:2002}).  The final processed lightcurves were downloaded from the Mikulski Archive for Space Telescopes (MAST) archive and included in our global fitting (see \S\ref{sec:GlobalModel}).

For our final transit fits, we adopt the SPOC 2-minute lightcurves when available but we re-extracted the 30-minute FFI light curves using a custom full frame image pipeline derived from that of \citet{Vanderburg:2019}. We downloaded the pixels surrounding the locations of each host star using the TESSCut interface \citep{Brasseur:2019} to the MAST. We first extracted light curves from a series of 20 different photometric apertures. We then removed systematic errors from each light curve by decorrelating with the mean and standard deviations of the spacecraft quaternion time series within each exposure and the \tess\ SPOC pipeline's Presearch Data Condition (PDC) cotrending basis vectors (binned to the cadence of each sector's observations). We performed the decorrelation via linear regression, where we solved for the best-fit coefficients for each model component using a matrix inversion technique, while iteratively excluding outlier points. We also included a basis spline in our linear regression model to simultaneously account for the stars' photometric variability. After subtracting the best-fit systematics components from our linear regression from the light curve, we then applied a correction for dilution from nearby stars customized for each of the 20 apertures based on a model of the \tess\ pixel response function and the known positions and magnitudes from the \tess\ Input Catalog (TIC, \citealp{Stassun:2018_TIC}) of nearby stars. Finally, for each star we selected one of the 20 photometric apertures by finding which one minimized its photometric scatter (outside of transit) and chose that as the final light curve for each star. We compared our final FFI lightcurve of TOI-2025 with that created by the SPOC pipeline and the MIT Quick Look Pipeline (QLP, \citealp{Huang:QLP}) as a check for the lightcurve quality (see Figure \ref{fig:LCcomp}). We adopt our custom FFI lightcurve for the final global fitting but note no significant difference in the transit properties when comparing the three versions of the FFI lightcurves. Additionally, we have photometric follow-up transits from the ground for each system other than TOI-2497.

To properly fit our \tess\ photometry within the global fit, we flatten the out-of-transit features using {\it Keplerspline}\footnote{\url{https://github.com/avanderburg/keplerspline}}, which fits a spline to the variability seen and divides out the best-fit model \citep{Vanderburg:2014}. The spline requires spacing for the break points  (breaks in the spline to handle discontinuities) and we optimized this by following the methodology from \citet{Shallue:2018} to  minimize the Bayesian information criterion. Most of the out-of-transit information provides little to no useful information in determining the full system parameters in the case of these six TOIs but is still computationally intensive to model. Therefore, we remove all baseline photometry from the \tess\ lightcurves, only keeping one full transit duration before the transit until one full transit duration after each transit. In the global model, we modeled all flattened lightcurve segments for each system of a given cadence with the same zero point and added variance (see \S\ref{sec:GlobalModel}). 

\subsection{KELT Photometry}
\label{sec:kelt}
Since \tess\ focuses on observing bright ($V$<12) stars, there is a wealth of archival data on these targets from even small-aperture surveys like the Kilodegree Extremely Little Telescope (KELT) survey\footnote{\url{https://keltsurvey.org}} \citep{Pepper:2007, Pepper:2012, Pepper:2018}. See \citet{Siverd:2012} \& \citet{Kuhn:2016} for a discussion on the KELT-North and KELT-South  observing strategy and reduction techniques. KELT uses two small aperture telescopes (Mamiya 645 80mm f/1.9 lens with 42mm aperture, Apogee 4k$\times$4k CCD) to observe most of the entire sky on a 20 to 30 minute cadence. Light curves from the KELT survey are accessible through the NASA Exoplanet Archive\footnote{\url{https://exoplanetarchive.ipac.caltech.edu/cgi-bin/TblSearch/nph-tblSearchInit?app=ExoTbls&config=kelttimeseries}}.

We do not recover the transits detected by \tess, likely due to a combination of the poor duty cycle from the ground (for the longer period systems, \citealp{Gaudi:2005}), the faintness of the host stars (for the shorter orbital period systems), and some of the transits being shallow ($<$0.5\%). However, KELT data can be useful to measure stellar rotation periods. Following the approach of \citet{Stassun:1999b, Oelkers:2018, Rodriguez:2021}, we executed a search for periodic signals using the KELT data. For these stars, we post-processed the light curve data using the Trend-Filtering Algorithm \citep{Kovacs:2005} to remove common systematics. We then searched for candidate rotation signals using a modified version of the Lomb-Scargle period finder algorithm \citep{Lomb:1976, Scargle:1982}. We searched for periods between a minimum period of 0.1 days and a maximum period of 100 days using the autopower feature of the \texttt{astropy} implementation of Lomb-Scargle. We masked periods between 0.5 and 0.505~days and 0.97--1.04~days to avoid the most common detector aliases associated with KELT's observational cadence and its interaction with the periods for the solar and sidereal day. For each star, we selected the highest statistically significant peak of the power spectrum as the candidate period for stellar variability. 

We then executed a boot-strap analysis, using 100 Monte-Carlo iterations, where the dates of the observations were not changed but the magnitude values of the light curve were randomized, following the work of \citet{Henderson:2012, VanderPlas:2018}. We recalculated the Lomb-Scargle power spectrum for each iteration, and recorded the maximum peak power of all iterations. If the highest power spectrum peak was larger than the maximum simulated peak after 100 iterations, we considered the periodic signal to be a candidate rotation period. We find only TOI-1811 to have a significant candidate rotation period at 25.779 days using KELT data. 

\subsection{WASP Photometry}
\label{sec:wasp}
Additional observations were available for only TOI-1811 from the Wide Angle Survey for Planets (WASP) survey. Each WASP site (La Palma and SAAO) used an array of eight 200-mm, f/1.8 lenses to create a large field of view \citep{Pollacco:2006}. The typical cadence of the observations were 15-30 minutes. Observations of TOI-1811 from 2007 and 2011 were available and following the techniques from \citet{Maxted:2011}, we searched for periodic modulation consistent with the rotation period of the star. We find a similar period to that what was in the KELT data, 23$\pm$1 days. Additionally, using the WASP search algorithm described in \citet{CollierCameron:2007} on the observations and the identification of planetary period of TOI-1811 b from \tess, we measure the WASP ephemeris of planet to be a period of 3.7130803$\pm$0.0000292 and a mid-transit epoch (T$_C$) of 2454006.04900$\pm$0.00337 HJD$_{\rm TDB}$. This ephemeris is consistent with the \tess\ ephemeris and therefore is used as a prior for the \texttt{EXOFASTv2} global analysis of TOI-1811 b (see \S\ref{sec:GlobalModel}).

\subsection{Ground-based Photometry from the {\it TESS} Follow-up Observing Program Working Group} \label{sec:sg1}
As part of the confirmation processes within TFOP, we observed five of the six  giant planet systems presented in this paper using a variety of small-aperture ($<$2 meter) telescopes to confirm the transit was on target and to refine the system parameters (particularly increasing the photometric baseline to improve our precision and accuracy on future times of transit). Observations were obtained using the Las Cumbres Observatory (LCO) telescope network \citep{Brown:2013},  KeplerCam on the 1.2m telescope at Fred Lawrence Whipple Observatory (FLWO), C. R. Chambliss Astronomical Observatory (CRCAO) at Kutztown University, Brigham Young University's campus telescopes, El Sauce Observatory, MUSCAT2 on the 1.5m Telescopio Carlos S\'{a}nchez (TCS) , the University of Louisville's Moore Observatory, Michigan State University's Observatory, George Mason University's Observatory, Optical Wide-field patroL network (OWL-Net) Oukaimeden observatory (OWL), Waffelow Creek Observatory, Observatori de Ca l'Ou, MASTER-Ural observatory, Villa '39 Observatory, Observatoire Priv\'e du Mont (OPM), Conti Private Observatory (CPO),  and Kotizarovci Observatory. Table \ref{tbl:detrending_parameters} shows the information on each observatory and the detrending parameters used within the global fit. The photometric observations were reduced and aperture photometry extraction was conducted using \texttt{AstroImageJ} \citep{Collins:2017} for all follow-up transit observations except MUSCAT2 and the MASTER-Ural observations. Below we briefly review the reduction process used for these facilities. Unfortunately, due to its longer orbital period, we were not able to get photometric follow-up on TOI-2497.

\begin{table*}
\scriptsize
\setlength{\tabcolsep}{2pt}
\centering
\caption{Literature and Measured Properties}
\begin{tabular}{llccccccccc}
  \hline
  \hline
Other identifiers\dotfill \\
& & TOI-1811 & TOI-2025 & TOI-2145 & TOI-2152 & TOI-2154 & TOI-2497 \\
& & TIC 376524552& TIC 394050135 & TIC 88992642 & TIC 395393265 & TIC 428787891 &  TIC 97568467 \\
& & --- & ---& HIP 86040 & --- & --- & HD 250208 \\
& TYCHO-2 & TYC 1992-00307-1 &TYC 4595-00797-1 & TYC 3091-00842-1 &TYC 4498-01400-1 & TYC 4617-00138-1 & TYC 0725-01745-1\\
&2MASS & J12354142+2712518& J18511077+8214436 &  J17350195+4041421 & J01452120+7747244&  J04440676+8421511& J06001500+1153030 \\ 
&TESS Sector & [22] & [14, 18, 19, 20, 24, 25, 26, 40] & [25, 26, 40] & [18, 19, 25, 26] & [19, 20, 25, 26] & [6, 33]\\
\hline
\hline
Parameter & Description & Value &Value &Value &Value &Value & Reference\\
\hline 
$\alpha_{J2000}\ddagger$\dotfill&Right Ascension (RA)\dotfill &12:35:41.419& 18:51:10.840 & 17:35:01.950 & 01:45:21.218 & 04:44:06.869 & 06:00:15.008 & 1 \\
$\delta_{J2000}\ddagger$\dotfill&Declination (Dec)\dotfill &+27:12:51.923& +82:14:43.562 & +40:41:42.205 & +77:47:24.623 & +84:21:51.119 & +11:53:03.031 & 1 \\
${\rm G}$\dotfill     & Gaia $G$ mag.\dotfill     & 11.76$\pm$0.02&11.36$\pm$0.02 & 8.94$\pm$0.02 & 11.24$\pm$0.02 & 11.04$\pm$0.02 & 9.47$\pm$0.02 & 1 \\
B$_{\rm P}$\dotfill			&Gaia B$_{\rm P}$ mag.\dotfill & 12.33$\pm$0.02&11.69$\pm$0.02 & 9.24$\pm$0.02 & 11.68$\pm$0.02 & 11.32$\pm$0.02 & 9.73$\pm$0.02 & 1 \\
R$_{\rm P}$\dotfill			&Gaia R$_{\rm P}$ mag.\dotfill & 11.07$\pm$0.02&10.90$\pm$0.03 & 8.52$\pm$0.02 & 10.65$\pm$0.02 & 10.61$\pm$0.02 & 9.10$\pm$0.02 & 1 \\
${\rm T}$\dotfill     & TESS mag.\dotfill     & 11.1237$\pm$0.0061&10.9461$\pm$0.0061 & 8.5594$\pm$0.0063 & 10.7053$\pm$0.0061 & 10.6611$\pm$0.0085 & 9.1411$\pm$0.0063 & 2 \\
&                \\
J\dotfill			& 2MASS J mag.\dotfill &10.280$\pm$0.024& 10.380$\pm$0.025 & 8.021$\pm$0.020 & 9.973$\pm$0.026 & 10.154$\pm$0.025 & 8.697$\pm$0.021 & 3 \\
H\dotfill			& 2MASS H mag.\dotfill & 9.732$\pm$0.027&10.071$\pm$0.028 & 7.810$\pm$0.023 & 9.669$\pm$0.30 & 9.864$\pm$0.027 & 8.533$\pm$0.020 & 3 \\
K$_S$\dotfill			& 2MASS ${\rm K_S}$ mag.\dotfill &9.643$\pm$0.025& 10.010$\pm$0.021 & 7.761$\pm$0.031 & 9.597$\pm$0.024 & 9.850$\pm$0.025 & 8.486$\pm$0.020 & 3 \\
&                \\
\textit{WISE1}\dotfill		& \textit{WISE1} mag.\dotfill &9.579$\pm$0.030& 9.995$\pm$0.030 & 7.706$\pm$0.030 & 9.535$\pm$0.030 &  9.808$\pm$0.030 & 8.418$\pm$0.030 & 4 \\
\textit{WISE2}\dotfill		& \textit{WISE2} mag.\dotfill &9.668$\pm$0.030& 10.037$\pm$0.030 & 7.745$\pm$0.030 & 9.543$\pm$0.030 & 9.836$\pm$0.030 & 8.448$\pm$0.030 & 4 \\
\textit{WISE3}\dotfill		& \textit{WISE3} mag.\dotfill &9.590$\pm$0.042& 9.973$\pm$0.042 & 7.717$\pm$0.030 & 9.470$\pm$0.032 & 9.771$\pm$0.038 & 8.424$\pm$0.030 & 4 \\
\textit{WISE4}\dotfill		& \textit{WISE4} mag.\dotfill & --- & --- & 7.691$\pm$0.122 & 8.968$\pm$0.304 & --- & 8.47$\pm$0.365 & 4 \\
&                \\
$\mu_{\alpha}$\dotfill		& Gaia DR2 proper motion\dotfill & -45.874$\pm$0.058&2.791$\pm$0.036 & -6.512$\pm$0.035 & 27.643$\pm$0.040 &	-10.783$\pm$0.036 &  12.502$\pm$0.076 & 1 \\	
                    & \hspace{3pt} in RA (mas yr$^{-1}$)	&&                \\
$\mu_{\delta}$\dotfill		& Gaia DR2 proper motion\dotfill 	& -10.766$\pm$0.035& -4.521$\pm$0.045 & -3.281$\pm$0.040 & -11.634$\pm$0.048 & 15.218$\pm$0.043&  -27.310$\pm$0.064& 1 \\
                    & \hspace{3pt} in DEC (mas yr$^{-1}$) &  &                \\
&                \\
$v\sin{i_\star}$\dotfill &  Rotational velocity (\kms) \hspace{9pt}\dotfill &3.3$\pm$0.5&7.3$\pm$0.5&19.4$\pm$0.5&5.4$\pm$0.5&5.4$\pm$0.5&39.6$\pm$1.0 & \S\ref{sec:TRES}\& \S\ref{sec:TRES}               \\
$\pi^\dagger$\dotfill & Gaia DR2 Parallax (mas) \dotfill &7.801$\pm$0.051&2.978$\pm$0.031&4.451$\pm$0.031&3.302$\pm$0.042&3.374$\pm$0.034&3.507$\pm$0.051&1\\
\\
\hline
\hline
\end{tabular}
\begin{flushleft}
 \footnotesize{ \textbf{\textsc{NOTES:}}
 The uncertainties of the photometry have a systematic error floor applied. \\
 $\ddagger$ RA and Dec are in epoch J2000. The coordinates come from Vizier where the Gaia RA and Dec have been precessed and corrected to J2000 from epoch J2015.5.\\
 $\dagger$ Values have been corrected for the -0.30 $\mu$as offset as reported by \citet{Lindegren:2018} but this is not significant for these systems.\\
 References are: $^1$\citet{Gaia:2018},$^2$\citet{Stassun:2018_TIC},$^3$\citet{Cutri:2003}, $^4$\citet{Cutri:2012}\\
}
\end{flushleft}
\label{tbl:LitProps}
\end{table*}

Two of our follow up transit observations did not use \texttt{AstroImageJ} to perform the reduction and photometry. TOI-1811 was observed on the night of UT 2021 June 05  with the multicolor imager MuSCAT2 \citep{Narita:2019} mounted on the 1.5 m Telescopio Carlos S\'{a}nchez (TCS) at Teide Observatory, Spain. The raw data were reduced by the MuSCAT2 pipeline \citep{Parviainen:2019} which performed a standard image calibration and aperture photometry. TOI-2152 was observed on UT 2020 December 12 with MASTER-Ural 0.4m telescope. The data reduction included standard dark, flat field and astrometry corrections, and is performed using the MASTER-Ural pipeline\footnote{\url{https://master.kourovka.ru/}}. Comparison stars were selected from the {\it Gaia} DR2 catalog. Aperture photometry of the object and the ensemble of comparison stars was performed using Python/Photutils \citep{Bradley:2019}. Photometric data processing and detrending was completed with the Python version of the Astrokit \citep{Burdanov:2014}, to minimize the standard deviation of the ensemble of comparison stars.


\subsection{Spectroscopy}
\label{sec:spectroscopy}
To confirm these six systems as bona fide transiting giant planets by removing any remaining false positive scenario, we obtained time-series spectroscopic measurements of each target coordinated through TFOP. These radial velocity measurements, combined with the transit photometry, allowed us to precisely measure the mass and orbital eccentricity of each system, a key component in understanding their evolutionary origins. Table \ref{tbl:rv} shows a sample radial velocity (RV) point per target per instrument (the full table will be available in machine-readable form in the online journal). The RVs and best-fit models from our \texttt{EXOFASTv2} analysis are shown in Figure \ref{fig:RVs} (see \S\ref{sec:GlobalModel}). 


\subsubsection{TRES Spectroscopy} \label{sec:TRES}
Using the Tillinghast Reflector Echelle Spectrograph \citep[TRES;][]{furesz:2008}\footnote{\url{http://www.sao.arizona.edu/html/FLWO/60/TRES/GABORthesis.pdf}} on the 1.5m Tillinghast Reflector, we measured the radial velocity orbit of all six TOIs presented in this paper. The telescope and spectrograph are located at the Fred L. Whipple Observatory (FLWO) on Mt. Hopkins, AZ. The reduction and RV analysis followed the procedure described in \citet{Buchhave:2010} and \citet{Quinn:2012}. The only difference is that the template spectra for the RV extraction were created by median-combining all of the out-of-transit spectra (after shifting each to align them). To rule out scenarios in which the apparent velocity variation is caused by a blended binary or stellar activity, we performed a line bisector analysis on the TRES spectra following the work of \citet{Torres:2007}. In all six cases, we find no evidence for these false positive scenarios in the bisector span variations. The bisector span measurements do not correlate with the derived radial velocities or orbital phase, and the measurements for each star agree to within the uncertainties. The TRES spectra were also analyzed using the Stellar Parameter Classification (SPC) package \citep{Buchhave:2012} to determine the \feh, \teff, and rotational velocity of each host star (see Tables \ref{tbl:LitProps} and \ref{tab:exofast_stellar}).

\subsubsection{CHIRON Spectroscopy} \label{sec:CHIRON}
We obtained 26 spectra of TOI-2497, between UT 2021 March 06 and UT 2022 March 25. The data were taken with the CHIRON \citep{Tokovinin:2013} high-resolution spectrograph, installed in the 1.5 m telescope at the Cerro Tololo International Observatory. The observations were performed with the image slicer (R $\sim$\,80000), with exposure times between 600s and 1800s, leading  to a SNR per extracted pixel between $\sim$\,20 - 80, at 550 nm. For each observation, we obtained a ThAr spectrum immediately before the science spectra to account for the instrument spectral drift, and thus a new  wavelength solution was automatically computed from that calibration, by the CHIRON pipeline \citep{Paredes:2021}.  The radial velocities were computed using an updated version of the pipeline used in \citet{Jones:2019}. A sample of the resulting values are listed in Table \ref{tab:rv}. 



\subsubsection{MINERVA Australis Spectroscopy} 
\label{sec:MINERVA}
We make use of the Minerva-Australis array for additional radial velocities of TOI-2497. Minerva-Australis is an array of four identical 0.7\,m telescopes located at Mt Kent Observatory, Australia. The telescopes are fed by four independent fibers into the KiwiSpec high resolution \'{e}chelle spectrograph, yielding a spectral resolving power of R$\sim$80,000 over the wavelength range of 5000-6300\r{A} \citep{Addison:2019}. Simultaneous wavelength calibration is provided by two calibration fibers, illuminated by a quartz lamp through an iodine cell, that tracks the instrument drift over an exposure. Radial velocities are measured from each telescope independently via a least-squares deconvolution between the extracted spectra and a synthetic, following the procedure described in \citet{Zhou:2021}. The template is generated from an ATLAS9 atmosphere model \citep{Castelli:2004} at the atmosphere parameters of the target star, and has no rotational broadening applied. The resulting line-broadening function is modeled with a kernel describing the rotational, macroturbulent, and instrumental broadening effects, as well as the radial velocity shift of a given exposure.

\begin{table*}
 \centering
 \caption{Photometric follow-up observations of these systems used in the global fits and the detrending parameters.}
 \label{tbl:detrending_parameters}
 \begin{tabular}{llllllllll}
    \hline
    \hline
Target & Observatory & Date (UT) & size (m) & Filter & FOV & Pixel Scale  & Exp (s) & Additive Detrending\\
    \hline

TOI-1811 b & LCO SSO & 2020 April 23 & 0.4 & $i^{\prime}$ & 19$\arcmin$ $\times$ 29$\arcmin$ &   0.57$\arcsec$ & 55 & airmass \\
TOI-1811 b & LCO HAL & 2020 April 23 & 0.4 & $i^{\prime}$ & 19$\arcmin$ $\times$ 29$\arcmin$ &   0.57$\arcsec$ & 55 & airmass \\
TOI-1811 b & BYU &  2020 April 27 & 0.6 & $R$ & 32$\arcmin$ $\times$ 32$\arcmin$ &  0.93$\arcsec$ & 70 & airmass \\
TOI-1811 b & ULMT &  2020 April 27 & 0.6 & $i^{\prime}$ & 26.8$\arcmin$ $\times$ 26.8$\arcmin$&  0.395$\arcsec$ & 128 & airmass \\
TOI-1811 b & Montcabrer &  2020 April 27 & 0.3 & $i^{\prime}$ & 45.8$\arcmin$ $\times$ 45.8$\arcmin$ & 0.9$\arcsec$ &120& airmass \\
TOI-1811 b & BYU-12 &  2020 May 08 & 0.6 & $V$ & 25$\arcmin$ $\times$ 19$\arcmin$ &  0.92$\arcsec$ & 90 & airmass \\
TOI-1811 b & El Sauce &  2020 May 12 & 0.36 & $B$ & 19$\arcmin$ $\times$ 13$\arcmin$ &  1.47$\arcsec$ & 180 & airmass \\
TOI-1811 b & LCO SSO &  2021 February 25  & 1.0 & $z^{\prime}$ & 27$\arcmin$ $\times$ 27$\arcmin$ &   0.39$\arcsec$ & 55 & airmass \\
TOI-1811 b & LCO SSO &  2021 February 25  & 1.0 & $B$ & 27$\arcmin$ $\times$ 27$\arcmin$ &   0.39$\arcsec$ & 70 & airmass \\
TOI-1811 b & MUSCAT2 &  2021 June 05  & 1.52 & $g^{\prime}$ &  7.4$\arcmin$ $\times$ 7.4$\arcmin$   &  0.44$\arcsec$ & 30 &  airmass\\
TOI-1811 b & MUSCAT2 &  2021 June 05  & 1.52 & $i^{\prime}$ & 7.4$\arcmin$ $\times$ 7.4$\arcmin$ &  0.44$\arcsec$  & 30 & airmass \\
TOI-1811 b & MUSCAT2 &  2021 June 05  & 1.52 & $r^{\prime}$ & 7.4$\arcmin$ $\times$ 7.4$\arcmin$ &   0.44$\arcsec$ & 15 & airmass \\
TOI-1811 b & MUSCAT2 &  2021 June 05  & 1.52 & $z^{\prime}$ &7.4$\arcmin$ $\times$ 7.4$\arcmin$  &  0.44$\arcsec$  & 30 & airmass \\ 

%
%
TOI-2025 b &Kotizarovci & 2020 June 26 & 0.3 & $TESS$ & 15$\arcmin$ $\times$ 23$\arcmin$ &   1.2064$\arcsec$ & 30 & airmass \\
TOI-2025 b &LCO TFN & 2020 June 26 & 0.4 & $i^{\prime}$ & 19$\arcmin$ $\times$ 29$\arcmin$ &   0.57$\arcsec$ & 60 & airmass \\
TOI-2025 b &LCO TFN & 2020 June 26 & 0.4 & $g^{\prime}$ & 19$\arcmin$ $\times$ 29$\arcmin$ &   0.57$\arcsec$ & 60 & airmass \\
TOI-2025 b &FLWO/KeplerCam &2021 May 12 & 1.2 &$B$ & 23.1$\arcmin$ $\times$ 23.1$\arcmin$&  0.672$\arcsec$ & 20 & airmass \\
TOI-2025 b &FLWO/KeplerCam &2021 May 12 & 1.2 &$i^{\prime}$ & 23.1$\arcmin$ $\times$ 23.1$\arcmin$&  0.672$\arcsec$ & 7 & airmass \\
TOI-2025 b &GMU & 2021 May 21 & 0.8 &$R$ & 23$\arcmin$ $\times$ 23$\arcmin$ &   0.34$\arcsec$ & 50 & airmass \\
TOI-2025 b &CRCAO & 2021 May 21 & 0.61 &$R$ & 19.5$\arcmin$ $\times$ 13$\arcmin$&  0.39$\arcsec$ & 120 & airmass \\
TOI-2025 b &MSU & 2021 September 30 & 0.61 &$Clear$ & 9.5$\arcmin$ $\times$ 9.5$\arcmin$&  0.55$\arcsec$ & 20 & None \\
TOI-2025 b &MSU & 2021 October 18 & 0.61 &$Clear$ & 9.5$\arcmin$ $\times$ 9.5$\arcmin$&  0.55$\arcsec$ & 30 & None \\
TOI-2025 b &CPO & 2021 December 29 & 0.61 &$V$ & 23$\arcmin$ $\times$ 18$\arcmin$&  1$\arcsec$ & 30 & total counts \\

TOI-2145 b & CRCAO & 2021 Sept 07 & 0.61 &$r^{\prime}$ &  19.5$\arcmin$ $\times$ 13$\arcmin$&  0.39$\arcsec$ & 20 & airmass \\
TOI-2145 b & Moore & 2021 Sept 07 & 0.61 &$i^{\prime}$ & 26.8$\arcmin$ $\times$ 26.8$\arcmin$&  0.39$\arcsec$ & 20 & airmass \\

TOI-2152 b &OWL& 2020 August 17 & 0.5 & $B$ & 1.1$^{\circ}$ $\times$1.1$^{\circ}$ & 1$\arcmin$ & 20 & airmass\\
TOI-2152 b &Waffelow Creek& 2020 October 11 & 0.36 & $g^{\prime}$ & 27$\arcmin$ $\times$15$\arcmin$   & 0.66$\arcsec$ & 90 & airmass\\
TOI-2152 b &Waffelow Creek& 2020 October 11 & 0.36 & $i^{\prime}$ &27$\arcmin$ $\times$15$\arcmin$& 0.66$\arcsec$ & 90 & airmass\\
TOI-2152 b &Observatori de Ca l'Ou& 2020 November 24 & 0.4 & $B$ & 19$\arcmin$ $\times$19$\arcmin$ & 1.14$\arcsec$ & 150 & airmass\\
TOI-2152 b &MASTER-Ural& 2020 December 12 & 0.4 & $R$ & 2$^{\circ}$ $\times$2$^{\circ}$  & 1.85$\arcsec$ & 80 & airmass\\
TOI-2152 b &CRCAO & 2021 June 28 & 0.61 &$R$ &  19.5$\arcmin$ $\times$ 13$\arcmin$&  0.39$\arcsec$ & 120 & airmass \\

TOI-2154 b & V39-0m4 & 2020 August 18 & 0.4 & $I$ & 32$\arcmin$ $\times$32$\arcmin$  & 0.95$\arcsec$ & 60 & airmass\\
TOI-2154 b & OPM & 2020 October 29 & 0.2 & $z^{\prime}$ &  39$\arcmin$ $\times$29$\arcmin$ & 0.69$\arcsec$ & 180 & airmass\\
TOI-2154 b & Observatori de Ca l'Ou & 2020 November 23 & 0.4 & $B$ &  19$\arcmin$ $\times$19$\arcmin$ & 1.14$\arcsec$ & 110 & airmass\\
TOI-2154 b & LCO McDonald &  2020 December 03  & 1.0 & $z^{\prime}$ & 27$\arcmin$ $\times$ 27$\arcmin$ &   0.39$\arcsec$ & 45 & airmass \\
TOI-2154 b &MSU & 2021 October 24 & 0.61 &$Clear$ & 9.5$\arcmin$ $\times$ 9.5$\arcmin$&  0.55$\arcsec$ & 60 & airmass \\

TOI-2497 b & None\\
\hline
 \end{tabular}
\begin{flushleft}
{\footnotesize \textbf{\textsc{NOTES:}} All the follow-up photometry presented in this paper is available in machine-readable form in the online journal. See \S D in the appendix of \citet{Collins:2017} for a description of each detrending parameter. 
}
\end{flushleft}
\end{table*}

\subsubsection{MINERVA North Spectroscopy} \label{sec:MINERVANorth}
The MINERVA North observations of TOI-2145 were made with the MINERVA telescope array and KiwiSpec Spectrograph \citep{Wilson:2019,Swift:2015}, which consists of four robotic telescopes at Whipple Observatory in Arizona, fiber fed to a temperature and pressure stabilized, R$\sim$80,000, iodine cell calibrated spectrograph. We obtained 24 observations with T1, 16 observations with T2, and 5 observations with T3 spanning from UT 2020 May 09 to UT 2021 May 31. We extracted 1D spectra from the 2D spectra with our standard methods.

The corresponding MINERVA RVs are computed from the 1D spectra with \texttt{pychell} using updated methods compared to those described in \cite{Cale:2019}. Each 1-dimensional spectrum is forward modeled on a per-order basis. The model accounts for the wavelength solution, instrumental profile (IP), continuum, tellurics, and stellar Doppler shift. An iodine vapor gas cell in the calibration unit constrains the wavelength solution and IP. We use the Fourier Transform Spectrometer (FTS) scan measured at NIST, described in \citet{Wilson:2019}. A synthetic BT-Settl model (T$_{\rm eff}$ = 6000 K, $\log g=3.5$, $(Fe/H)_{\odot}=0$) is used as an initial stellar template, which is further Doppler broadened to $v \sin i = 19$ km\,s$^{-1}$ with PyAstronomy \citep{Czesla:2019}. \texttt{pychell} then iteratively updates this template based on the residuals between the data and model, and although the fits suggest the stellar template is more accurate at later iterations, the corresponding RVs are inconsistent with the orbit of the planet, whereas the initial BT-Settl template yields consistent RVs with the TRES observations which strongly support the planetary orbit. To ensure that the MINERVA North observations were not improperly influencing our results, we ran a global fit using only the TRES RVs and the results were consistent to 1$\sigma$. We have yet to find cause for the loss of accuracy at later iterations, and is a subject of future work. We therefore use RVs from the first iteration. The RMS of the residuals of our adopted RV model suggest a median $S/N$ per-spectral pixel of 17.

\begin{figure*}
	\centering\vspace{.0in}
	\includegraphics[width=0.33\linewidth, trim={2.5cm 13.0cm 9.4cm 8.5cm}, clip]{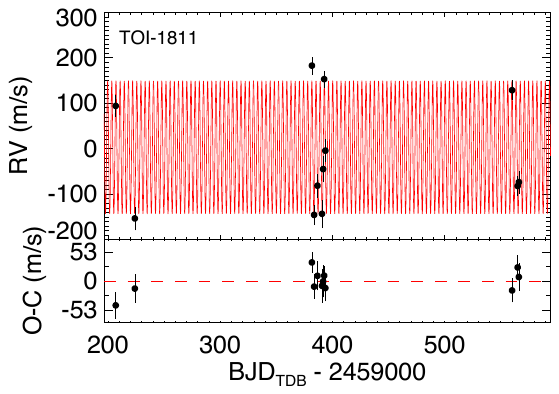}\includegraphics[width=0.33\linewidth, trim={2.5cm 13.0cm 9.4cm 8.5cm}, clip]{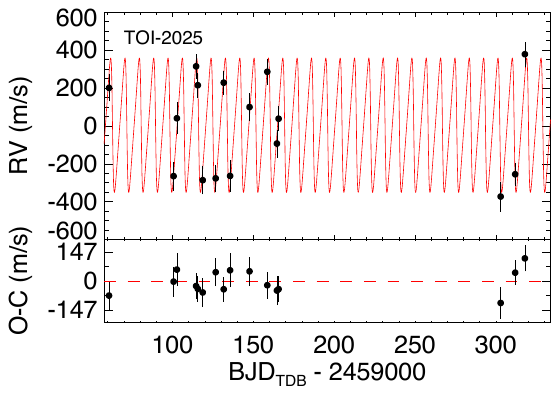}\includegraphics[width=0.33\linewidth, trim={2.5cm 13.0cm 9.4cm 8.5cm}, clip]{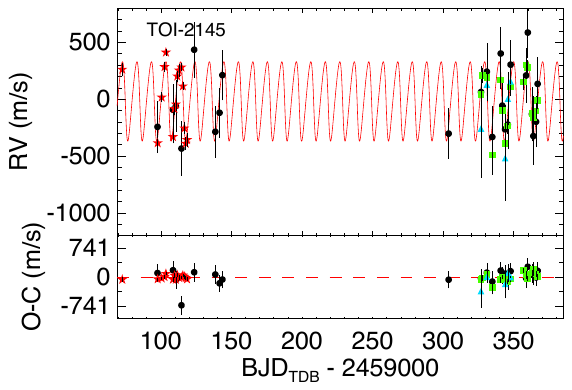}
	\includegraphics[width=0.33\linewidth, trim={2.5cm 13.0cm 9.4cm 8.5cm}, clip]{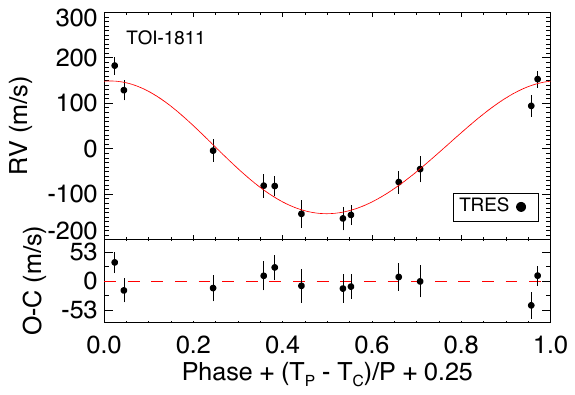}\includegraphics[width=0.33\linewidth, trim={2.5cm 13.0cm 9.4cm 8.5cm}, clip]{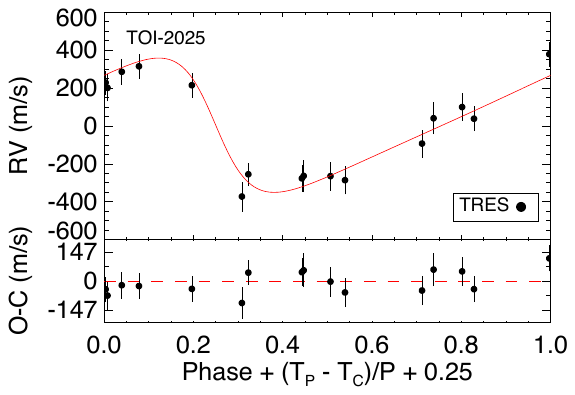}\includegraphics[width=0.33\linewidth, trim={2.5cm 13.0cm 9.4cm 8.5cm}, clip]{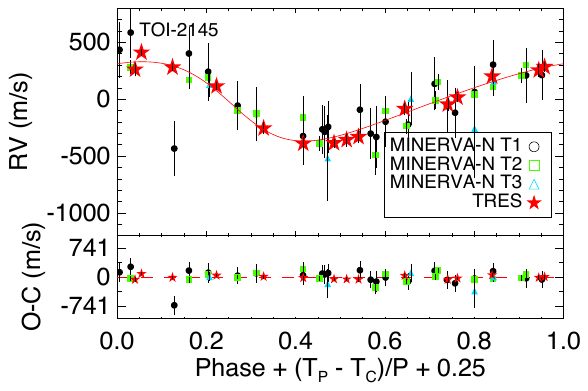}\\
	\includegraphics[width=0.33\linewidth, trim={2.5cm 13.0cm 9.4cm 8.5cm}, clip]{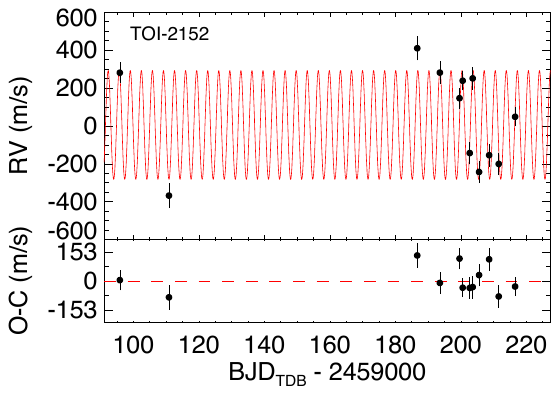}\includegraphics[width=0.33\linewidth, trim={2.5cm 13.0cm 9.4cm 8.5cm}, clip]{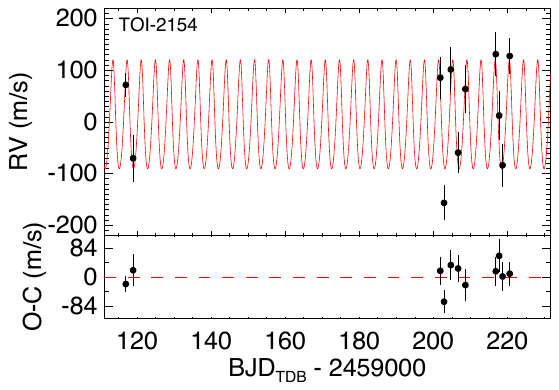}\includegraphics[width=0.33\linewidth, trim={2.5cm 13.0cm 9.4cm 8.5cm}, clip]{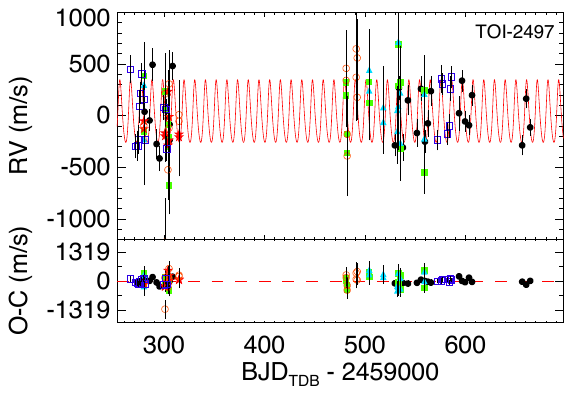}
	\includegraphics[width=0.33\linewidth, trim={2.5cm 13.0cm 9.4cm 8.5cm}, clip]{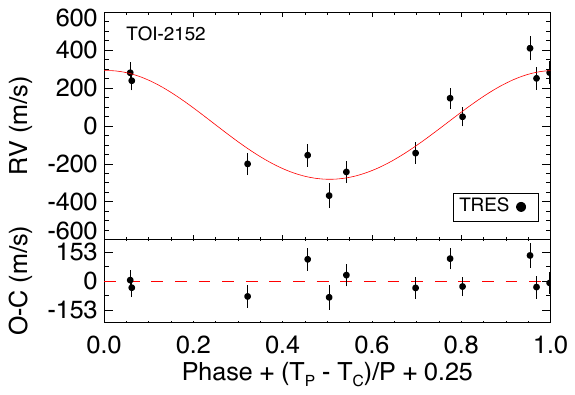}\includegraphics[width=0.33\linewidth, trim={2.5cm 13.0cm 9.4cm 8.5cm}, clip]{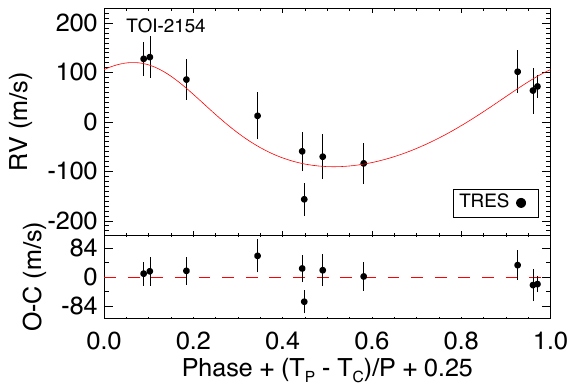}\includegraphics[width=0.33\linewidth, trim={2.5cm 13.0cm 9.4cm 8.5cm}, clip]{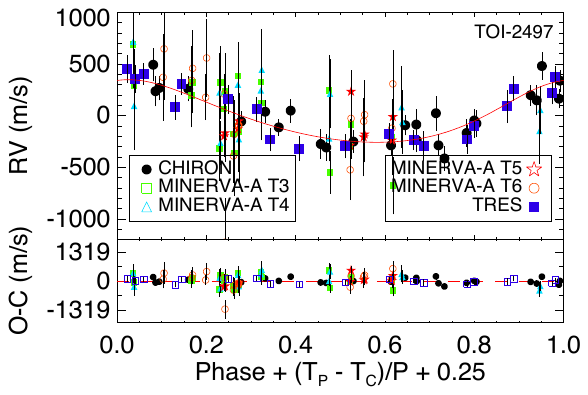}
	\caption{The RV observations of TOI-1811 (top-left), TOI-2025 (top-middle), TOI-2145 (top-right), TOI-2152 (bottom-left), TOI-2154 (bottom-middle), and TOI-2497 (bottom-right). In each case, the top figure shows the RVs vs time and the bottom panel is phased to the best-fit ephemeris from our global fit. The \texttt{EXOFASTv2} model is shown in red and the residuals to the best-fit are shown below each plot.}
	\label{fig:RVs} 
\end{figure*}

\begin{table}
\scriptsize
 \centering
 \caption{One RV point from each spectrograph for all six systems. The full table of RVs for each system is available in machine-readable form in the online journal. \label{tab:rv}}
 \label{tbl:rv}
 \begin{tabular}{llllllllll}
    \hline
    \hline
\bjdtdb & RV (m s$^{-1}$) & $\sigma_{RV}^{\dagger}$ (m s$^{-1}$) & Target & Instrument\\
\hline
2459206.966936 & -91.2 & 24.7 & TOI-1811 & TRES\\
2459060.767111 & 395.0 & 39.8 & TOI-2025 & TRES\\
2459097.65584 & -194.7 & 103.1& TOI-2145 & MINERVA T1\\
2459326.76336 & 311.8 & 51.8 & TOI-2145 & MINERVA T2\\
2459330.93081 & 94.6 & 167.0& TOI-2145 & MINERVA T3\\
2459072.693111 & 70.6 & 60.6 & TOI-2145 & TRES\\
2459095.870551 & 481.26 & 40.01 & TOI-2152 & TRES\\
2459201.888606 & 90.4 & 43.7 & TOI-2154 & TRES\\
2459596.60687 &385.0 &76.0 & TOI-2497 & CHIRON \\
2459279.908926 & 56275.1 & 271.6 & TOI-2497 & M-Australis T3\\
2459504.223682 & 56250.9 & 316.1 & TOI-2497 & M-Australis T4\\
2459279.908926 & 56019.7 & 340.9 & TOI-2497 & M-Australis T6\\
2459271.812986 & -634.1 & 71.0 & TOI-2497 & TRES\\
\hline
 \end{tabular}
\begin{flushleft}
{\footnotesize \textbf{\textsc{NOTES:}}}$^{\dagger}$ The internal RV error for the observation shown.  
\end{flushleft}
\end{table}

\subsection{High Resolution Imaging}
As part of our standard process for validating transiting exoplanets to assess the possible contamination of bound or unbound companions on the derived planetary radii \citep{Ciardi:2015}, we observed the TOIs with a combination of high-resolution imaging resources including near-infrared adaptive optics (AO) imaging at Lick (TOI-2145, TOI-2497) and Palomar (TOI-1811, TOI-2145) Observatories and with optical speckle imaging using the 2.5m SAI telescope (TOI-1811, TOI-2025, TOI-2145, TOI-2152, TOI-2154) and the Southern Astrophysical Research (SOAR) telescope (TOI-2497).  While the optical speckle observations tend to provide higher resolution, the NIR AO obsevations tend to provide better sensitivity, especially to lower-mass stars. If a companion is detected, the combination of the observations in multiple filters enables better characterization. Additionally, recent studies have shown that Gaia (DR2 and eDR3) \citep{Gaia:2018} is most efficient at identifying companions with separations greater than $\sim 0.5-1\arcsec$ \citep{Ziegler:2018}. Gaia eDR3 \citep{Gaia:2021} is also used to identify targets that have a large Renormalised Unit Weight Error (RUWE) value indicative of a poor astrometric fit assuming a single-star model and possibly indicating the presence of undetected stellar companions. For all of the observations, We only detect one faint companion to TOI-2152 ($\Delta$Mag $\sim$ 5) within 1\arcsec\ of the primary target. 
    
\subsubsection{Summary of AO Observations}
The Palomar Observatory observations of TOI-1811 and TOI-2145 were made with the PHARO instrument \citep{Hayward:2001} behind the natural guide star AO system P3K \citep{Dekany:2013} on UT 2021 February 23 and UT 2021 February 24, respectively, in a standard 5-point quincunx dither pattern with steps of 5\arcsec\ in the narrow-band $Br-\gamma$ filter $(\lambda_o = 2.1686; \Delta\lambda = 0.0326~\mu$m).  Each dither position was observed three times, offset in position from each other by 0.5\arcsec\ for a total of 15 frames; with an integration time of 30 and 1.4 seconds per frame, respectively for total on-source times of 450 and 21 seconds. PHARO has a pixel scale of $0.025\arcsec$ per pixel for a total field of view of $\sim25\arcsec$.

We also observed TIC 88992642 (TOI-2145) and TIC 97568467 (TOI-2497) on UT 2021 March 29 using the ShARCS camera on the Shane 3-meter telescope at Lick Observatory \citep{Kupke:2012, Gavel:2014, McGurk:2014}. Observations were taken with the Shane adaptive optics system in natural guide star mode in order to search for nearby, unresolved stellar companions. For each target, we collected sequences of observations using a $Ks$ filter ($\lambda_0 = 2.150$ $\mu$m, $\Delta \lambda = 0.320$ $\mu$m) and a $J$ filter ($\lambda_0 = 1.238$ $\mu$m, $\Delta \lambda = 0.271$ $\mu$m). We reduced the data using the publicly available \texttt{SImMER} pipeline \citep{Savel:2020}.\footnote{\url{https://github.com/arjunsavel/SImMER}} We find no nearby stellar companions within our detection limits.

The AO data were processed and analyzed with a custom set of IDL tools.  The science frames were flat-fielded and sky-subtracted.  The flat fields were generated from a median average of dark subtracted flats taken on-sky.  The flats were normalized such that the median value of the flats is unity.  The sky frames were generated from the median average of the 15 dithered science frames; each science image was then sky-subtracted and flat-fielded.  The reduced science frames were combined into a single combined image using an intra-pixel interpolation that conserves flux, shifts the individual dithered frames by the appropriate fractional pixels, and median-coadds the frames.  The final resolutions of the combined dithers were determined from the FWHM of the point spread functions for each of the stars: 0.102\arcsec\ for TOI-1811 and 0.092\arcsec\ for TOI-2145.  The sensitivities of the final combined AO image were determined by injecting simulated sources azimuthally around the primary target every $20^\circ $ at separations of integer multiples of the central source's FWHM \citep{Furlan:2017}. The brightness of each injected source was scaled until standard aperture photometry detected it with $5\sigma $ significance. The resulting brightness of the injected sources relative to primary target set the contrast limits at that injection location. The final $5\sigma$ limit at each separation was determined from the average of limits at that separation (across all azimuthal samples) and the uncertainty on the limit was set by the rms dispersion of the azimuthal slices at a given radial distance.  For both TOI-1811 and TO-2145, no additional stellar companions were detected in agreement with the other observations.

\subsubsection{Speckle Imaging}
\label{sec:SPECKLE}
Using the 4.1-m SOAR telescope, we obtained speckle imaging of TOI-2497 using HR Cam on UT 2021 February 27 in the $I$-band following the observing and reduction strategy described in \citet{Tokovinin:2018}. HRCam on SOAR has a 15$\arcsec\times$15$\arcsec$ field of view and had a 0.01575$\arcsec$ pixel scale. With a contrast of $\Delta$Mag of 7.7 at 1$\arcsec$, we detected no nearby companions around TOI-2497. For a complete description of the observing strategy for \tess\ targets, see \citet{Ziegler:2020}.

TOI-1811, TOI-2025, TOI-2145, TOI-2152, and TOI-2154 were observed with the Speckle Polarimeter \citep{Safonov:2017} on the 2.5~m telescope at the Caucasian Observatory of Sternberg Astronomical Institute (SAI) of Lomonosov Moscow State University. SPP uses Electron Multiplying CCD Andor iXon 897 as a detector. The atmospheric dispersion compensator allowed observation of relatively faint targets through the wide-band $I_c$ filter. For TOI-2145 we used a medium band interference filter with FWHM of 50~nm and centered on 625~nm. The power spectrum was estimated from 4000 frames with 30~ms exposure. The detector has a pixel scale of 20.6 mas pixel$^{-1}$. For all targets except for TOI-2152 we did not detect stellar companions, the contrast limits at 1$\arcsec$ are $\Delta$mag = 6.7 (TOI-1811), 6.4 (TOI-2025), 3.3 (TOI-2145), 5.9 (TOI-2152, this had multiple observations ranging from 4.7 to 6.3), and 6.5 (TOI-2154). We note that the difference image analysis performed in the data validation reports from \TESS\ show that the source of the transit signal for TOI-2145 was located within 5.0$\pm$2.7$\arcsec$ and for TOI-1811 was within 1.78$\pm$2.5$\arcsec$, complementing the high resolution imaging results. 

TOI-2152 is the only star that we found to have a close-in stellar companion. The separation, position, and contrast of the TOI-2152 inner companion were estimated on 4 dates; the results are presented in Table~\ref{table:sppobsTOI2152}. According to proper motion from Gaia eDR3, the primary star is expected to move by $22\pm0.02$~mas  over the period of our observations, from UT 2020 October 21 to UT 2021 July 17; however, there apparent motion is only $13\pm 11$ mas which is consistent with no discernible separation change. While not definitive, the companion appears to be a common proper motion companion and is likely gravitationally bound.  With a contrast of $\Delta I = 4.8$ mag, the detection is consistent with the companion being an M1V star \citep[($M\sim0.5 M_\odot$; $T_{eff}\sim 3600 K$; ][]{PM2013}.  At a distance of $\sim 320$pc, the companion has a projected separation of $\sim 250$au.  Interestingly, TOI-2152 also has another companion further out detected by Gaia with an angular separation of $\sim20$\arcsec\ ($\sim 6000$au; see \S\ref{gaia-assessment}).

\subsection{Gaia Assessment}\label{gaia-assessment}
In addition to the high resolution imaging, we have utilized Gaia to identify any\ wide stellar companions that may be bound members of the system.  Typically, these stars are already in the \tess\ Input Catalog and their flux dilution to the transit has already been accounted for in the transit fits and associated derived parameters.  Based upon similar parallaxes and proper motions \citep{Mugrauer:2020,Mugrauer:2021}, the only TOI in our sample which appears to have a wide stellar companion is TOI-2152 (in addition to the close-in companion identified in \S\ref{sec:SPECKLE}); the wide companion TIC~395393263 (Gaia~DR3~562112709676597376) is 20\arcsec\ to the NW ($PA \approx 300^\circ$) which corresponds to a projected physical separation of $\sim 6000 au$.  The companion has a mass and temperature consistent with an M4V star ($M\sim0.24 M_\odot$; $T_{eff}\sim 3223 K$ \citealp{Mugrauer:2021}) -- for such a small star at such a large separation, the stellar companion does not affect the stability of the planets or the measured radial velocities.  Interestingly, the projected positions on the sky of the three stars are not in a line indicating that the mutual inclination of the two stellar companions is non-zero - astrometric and/or radial velocity observations would be needed to determine if the transiting planet is aligned or not with either of the two stellar companions. A summary of the hierarchical triple TOI-2152 is given in Table~\ref{table:stellarTOI2152}. 
    
Gaia DR3 astrometry \citep{Gaia:2021} provides additional information on the possibility of inner companions that may have gone undetected by either Gaia DR2 data or the high resolution imaging. The Gaia Renormalised Unit Weight Error (RUWE) is a metric, similar to a reduced chi-square, where values that are $\lesssim 1.4$  indicate that the Gaia astrometric solution is consistent with the star being single whereas RUWE values $\gtrsim 1.4$ may indicate an astrometric excess noise, possibly caused the presence of an unseen companion \citep[e.g., ][]{Ziegler:2020}.  All of the TOIs in this sample, except TOI~1811, have RUWE values of $<1.1$ indicating that the astrometric fits are consistent with the single star model.  The RUWE for TOI-1811 is 1.66; there is no clear fixed boundary for when the RUWE unambiguously identifies the presence of an unseen stellar companion.  The transit of TOI-1811 is very deep (19 mmag in the \tess\ light curves) and with a short orbital period of 3.7 days, it may be the transit of the planet itself that is affecting the Gaia RUWE value.

\begin{table}
\caption{Binarity parameters of TOI-2152B on the basis of SPP observations: separation, position angle and magnitude difference in $I$ band.
\label{table:sppobsTOI2152}}
\begin{center}
\begin{tabular}[t]{|l|l|l|l|}
\hline
Date (UT)       &    $\rho^{\prime\prime}$ & P.A.$^{\circ}$& $\Delta$m \\
\hline
2020 Oct 21 &    $0.765\pm0.008$ &     $85.2\pm0.2$ &  $4.8\pm0.2$ \\
2020 Oct 28 &    $0.762\pm0.009$ &     $86.1\pm0.3$ &  $4.8\pm0.1$ \\
2020 Dec 02 &    $0.770\pm0.008$ &     $87.0\pm0.2$ &  $4.8\pm0.1$ \\
2021 Jul 17 &    $0.782\pm0.008$ &     $85.8\pm0.2$ &  $4.6\pm0.1$ \\
\hline
\end{tabular}
\begin{flushleft}
{\footnotesize \textbf{\textsc{NOTES:}}} The $\rho,^{\prime\prime}$ is the projected separation of the neighbor, if at the distance of the primary star.   
\end{flushleft}
\end{center}
\end{table}

\begin{table*}
\begin{center}
\caption{Estimated Parameters for TOI-2152 Stellar Components
\label{table:stellarTOI2152}}
\begin{tabular}[p]{l|c|c|c|c|c|c}
\hline
Stellar  & Separation & Mass & Radius & T$_{eff}$ & Spectral & Notes\\
Component & [au] & [$M_\odot$] & [$R_\odot$] & [K] & Type\\
\hline
TOI-2152A &    $\cdots$ & 1.52 & 1.61 & 6630 & F4V & Table~\ref{tab:exofast_stellar}\\
TOI-2152B &    250 & 0.5 & 0.4 & 3600 & M1V &\cite{Pecaut:2013, boyajian2012}\\
TOI-2152C &    6000 & 0.24 & 0.2 & 3200 & M4V &\cite{Mugrauer:2020, boyajian2012}\\
\hline
\end{tabular}
\end{center}
\end{table*}


\section{EXOFAST\lowercase{v}2 Global Fits} 
\label{sec:GlobalModel}
Following the same strategy laid out in \S3 of \citet{Rodriguez:2021}, we globally fit the RVs, \TESS\ and TFOP photometry (see Figures \ref{fig:transits1}, \ref{fig:transits2}, \& \ref{fig:RVs}; and  \S\ref{sec:Obs}) for TOI-1811 b, TOI-2025 b, TOI-2145 b, TOI-2152A b, TOI-2154 b, and TOI-2497 b with \texttt{EXOFASTv2} \citep{Eastman:2013, Eastman:2019} to determine their individual system parameters and place them in context with the known exoplanet population. To ensure that none of the SG1 partial transits are influencing the results for any system, we run fits with and without partial transits and the fitted system parameters are consistent ($<1\sigma$). The SG1 photometry provides a strong constrain on the transit ephemerides of these systems by significantly extending the baseline of the observations. The Spectral Energy Distribution (SED) and the MESA Isochrones and Stellar Tracks (MIST) stellar evolution models \citep{Paxton:2011, Paxton:2013, Paxton:2015, Choi:2016, Dotter:2016} were included to constrain the host star's parameters within the fit, and we account for the 30 minute smearing in the \tess\ FFI lightcurves. We enforced a systematic limit on the precision broad-band photometry (see  Table \ref{tbl:LitProps}, \citealp{Stassun:2016}) and use \texttt{EXOFASTv2}'s default lower limit on the systematic error on the bolometric flux (F$_{\rm bol}\sim$3\%). We adopted a Gaussian prior on the \feh, parallax from Gaia DR2 (\citealp{Gaia:2016, Gaia:2018}, correcting for the offset reported by \citealp{Lindegren:2018}), and an upper bound on the line of sight extinction from \citet{Schlegel:1998} \& \citet{Schlafly:2011}. Both SPOC and our custom pipeline correct the \tess\ photometry for known nearby blended stars in the aperture. To allow some flexibility while checking this correction, we fit for dilution term on the \tess\ band, and placed a Gaussian prior of $0\pm$10\% of the contamination ratio reported by the \TESS\ Input Catalog (TIC,  \citealp{Stassun:2018_TIC}). We saw no evidence of any significant dilution in TOI-1811, TOI-2025, TOI-2152, and TOI-2154. Unfortunately, without an independent full transit for TOI-2145 and TOI-2497, we are not able to perform this test with the limited amount of photometric follow-up. We use the recommended convergence criteria by \citet{Eastman:2019} of a Gelman-Rubin statistic ($<$1.01) and independent draws ($>$1000). The results for each system are in Tables \ref{tab:exofast_stellar}, \ref{tab:exofast_other1}, \& \ref{tab:exofast_other2} and in Figures \ref{fig:transits1}, \ref{fig:transits2}, \& \ref{fig:RVs}.

\begin{table*}
\tiny
\centering
\caption{Median values and 68\% confidence intervals for the global models}
\begin{tabular}{llccccccc}
  \hline
  \hline
  \\
\multicolumn{2}{l}{Priors:}& TOI-1811 b &TOI-2025 b &TOI-2145 b & TOI-2152A b & TOI-2154 b & TOI-2497 b\\
\hline\\
Gaussian & $\pi$ Gaia Parallax (mas) \dotfill & 7.800$\pm$0.051 &2.9775$\pm$0.0312&4.4509$\pm$0.0314&3.3018$\pm$0.0417&3.3744$\pm$0.0337&3.5072$\pm$0.0508\\
Gaussian & $[{\rm Fe/H}]$ Metallicity (dex) & 0.27$\pm$0.08 & 0.19$\pm$0.08 & 0.23$\pm$0.08 & 0.27$\pm$0.08 & 0.02$\pm$0.08& 0.06$\pm$0.08\\
Upper Limit & $A_V$ V-band extinction (mag) & 0.04619 & 0.1801 & 0.1004 & 2.0150 &0.2793&1.3020\\
Gaussian & $T_C^{**}$ Time of conjunction (HJD$_{\rm TDB}$)\dotfill &2454006.04900$\pm$0.00337&---& ---&---&---&---\\
Gaussian$^{\prime}$& $D_T$ Dilution in \tess\  \dotfill &0.00000$\pm$0.00028&0.00000$\pm$0.00026&---&0.00000$\pm$0.02032&0.00000$\pm$0.00055&---\\
  \hline
  \hline
Parameter & Units & Values &Values & Values&Values&Values&Values \\
\multicolumn{2}{l}{Stellar Parameters:}&&&&\\
~~~~$M_*$\dotfill &Mass (\msun)\dotfill & $0.817^{+0.033}_{-0.030}$& $1.199^{+0.074}_{-0.092}$& $1.720^{+0.057}_{-0.075}$& $1.516^{+0.085}_{-0.10}$& $1.233^{+0.077}_{-0.090}$& $1.859^{+0.087}_{-0.083}$\\
~~~~$R_*$\dotfill &Radius (\rsun)\dotfill & $0.769^{+0.019}_{-0.017}$& $1.459^{+0.043}_{-0.042}$& $2.749^{+0.066}_{-0.064}$& $1.612^{+0.056}_{-0.051}$& $1.396^{+0.049}_{-0.043}$& $2.36^{+0.12}_{-0.11}$\\
~~~~$R_{*,SED}$\dotfill &Radius (\rsun)\dotfill & $0.7717^{+0.0074}_{-0.0073}$& $1.498\pm0.018$& $2.792^{+0.029}_{-0.028}$& $1.621^{+0.038}_{-0.036}$& $1.400\pm0.020$& $2.365\pm0.046$\\
~~~~$L_*$\dotfill &Luminosity (\lsun)\dotfill & $0.2753^{+0.0071}_{-0.0070}$& $2.44^{+0.12}_{-0.11}$& $9.92\pm0.33$& $4.50^{+0.77}_{-0.64}$& $2.72^{+0.23}_{-0.20}$& $14.7^{+2.0}_{-1.7}$\\
~~~~$F_{Bol}$\dotfill &Bolometric Flux$\times$10$^{-9}$ (cgs)\dotfill & $0.536\pm0.012$& $0.693^{+0.033}_{-0.028}$& $6.29^{+0.18}_{-0.28}$& $1.57^{+0.26}_{-0.22}$& $0.992^{+0.080}_{-0.071}$& $5.77^{+0.76}_{-0.65}$\\
~~~~$\rho_*$\dotfill &Density (cgs)\dotfill & $2.54^{+0.17}_{-0.19}$& $0.542^{+0.062}_{-0.058}$& $0.1161^{+0.0090}_{-0.0091}$& $0.511^{+0.062}_{-0.066}$& $0.639^{+0.083}_{-0.087}$& $0.200^{+0.034}_{-0.029}$\\
~~~$\log{g}$\dotfill &Surface gravity (cgs)\dotfill & $4.579^{+0.022}_{-0.025}$& $4.187^{+0.037}_{-0.040}$& $3.794^{+0.023}_{-0.027}$& $4.205^{+0.038}_{-0.048}$& $4.239^{+0.042}_{-0.051}$& $3.962^{+0.050}_{-0.049}$\\
~~~~$T_{\rm eff}$\dotfill &Effective Temperature (K)\dotfill & $4766^{+52}_{-54}$& $5977^{+79}_{-78}$& $6177\pm67$& $6630^{+300}_{-290}$& $6280\pm160$& $7360^{+320}_{-300}$\\
~~~~$T_{\rm eff,SED}$\dotfill &Effective Temperature (K)\dotfill & $4760^{+20}_{-19}$& $5896^{+81}_{-69}$& $6134^{+42}_{-53}$& $6610^{+320}_{-300}$& $6270^{+150}_{-140}$& $7350^{+270}_{-250}$\\
~~~~$[{\rm Fe/H}]$\dotfill &Metallicity (dex)\dotfill & $0.306^{+0.076}_{-0.077}$& $0.15^{+0.11}_{-0.10}$& $0.247^{+0.074}_{-0.072}$& $0.282^{+0.075}_{-0.079}$& $0.011^{+0.071}_{-0.059}$& $0.094^{+0.076}_{-0.072}$\\
~~~~$[{\rm Fe/H}]_{0}^\dagger$\dotfill &Initial Metallicity \dotfill & $0.280^{+0.074}_{-0.076}$& $0.200^{+0.095}_{-0.088}$& $0.286^{+0.070}_{-0.066}$& $0.368^{+0.063}_{-0.070}$& $0.105^{+0.059}_{-0.056}$& $0.177^{+0.075}_{-0.074}$\\
~~~$Age$\dotfill &Age (Gyr)\dotfill & $5.9^{+4.9}_{-4.0}$& $4.5^{+2.3}_{-1.5}$& $1.80^{+0.33}_{-0.23}$& $0.83^{+1.1}_{-0.58}$& $2.9^{+2.1}_{-1.5}$& $1.00^{+0.22}_{-0.19}$\\
~~~~$EEP^\ddagger$\dotfill &Equal Evolutionary Phase \dotfill & $335^{+15}_{-33}$& $407^{+29}_{-35}$& $402.2^{+11}_{-8.7}$& $328^{+22}_{-35}$& $369^{+46}_{-33}$& $360.4^{+12}_{-9.7}$\\
~~~~$A_V$\dotfill &V-band extinction (mag)\dotfill & $0.024^{+0.015}_{-0.016}$& $0.079^{+0.058}_{-0.050}$& $0.071^{+0.022}_{-0.035}$& $0.98^{+0.17}_{-0.18}$& $0.137^{+0.088}_{-0.087}$& $0.52\pm0.13$\\
~~~~$\sigma_{SED}$\dotfill &SED photometry error scaling & $0.86^{+0.35}_{-0.21}$& $0.50^{+0.21}_{-0.13}$& $0.92^{+0.33}_{-0.21}$& $1.09^{+0.40}_{-0.25}$& $0.67^{+0.27}_{-0.16}$& $0.70^{+0.28}_{-0.17}$\\
~~~~$\varpi$\dotfill &Parallax (mas)\dotfill & $7.800^{+0.050}_{-0.051}$& $2.978\pm0.031$& $4.451\pm0.031$& $3.302\pm0.042$& $3.374\pm0.034$& $3.507^{+0.050}_{-0.051}$\\
~~~~$d$\dotfill &Distance (pc)\dotfill & $128.21^{+0.84}_{-0.82}$& $335.8\pm3.5$& $224.7\pm1.6$& $302.8^{+3.9}_{-3.8}$& $296.3^{+3.0}_{-2.9}$& $285.1^{+4.2}_{-4.0}$\\
\multicolumn{2}{l}{Planetary Parameters:}&&&&\\
~~~~$P$\dotfill &Period (days)\dotfill & $3.7130765\pm0.0000017$& $8.8720982\pm0.0000077$& $10.26075^{+0.00083}_{-0.00082}$& $3.3773512^{+0.0000060}_{-0.0000061}$& $3.8240801\pm0.0000025$& $10.655669\pm0.000038$\\
~~~~$R_P$\dotfill &Radius (\rj)\dotfill & $0.994^{+0.025}_{-0.023}$& $1.061^{+0.033}_{-0.031}$& $1.069^{+0.029}_{-0.028}$& $1.281^{+0.050}_{-0.046}$& $1.453^{+0.053}_{-0.048}$& $0.994^{+0.055}_{-0.049}$\\
~~~$M_P$\dotfill &Mass (\mj)\dotfill & $0.972^{+0.076}_{-0.078}$& $3.60\pm0.33$& $5.26^{+0.38}_{-0.37}$& $2.83^{+0.38}_{-0.37}$& $0.92^{+0.19}_{-0.18}$& $4.82\pm0.41$\\
~~~~$T_C$\dotfill &Time of conjunction (\bjdtdb)\dotfill & $2458899.87080^{+0.00019}_{-0.00018}$& $2458690.28895\pm0.00043$& $2459013.2808\pm0.0011$& $2458792.55575\pm0.00034$& $2458819.73080^{+0.0011}_{-0.00079}$& $2459205.0992^{+0.0011}_{-0.0010}$\\
~~~~$T_0^\star$\dotfill &Optimal conjunction Time (\bjdtdb)\dotfill & $2459126.36846^{+0.00017}_{-0.00014}$& $2459089.53337\pm0.00025$& $2459023.54156^{+0.00071}_{-0.00070}$& $2458927.64980\pm0.00023$& $2459148.60166^{+0.0011}_{-0.00072}$& $2459109.19818^{+0.0011}_{-0.00100}$\\
~~~~$a$\dotfill &Semi-major axis (AU)\dotfill & $0.04389^{+0.00058}_{-0.00055}$& $0.0892^{+0.0018}_{-0.0023}$& $0.1108^{+0.0012}_{-0.0016}$& $0.05064^{+0.00093}_{-0.0011}$& $0.0513^{+0.0011}_{-0.0013}$& $0.1166\pm0.0018$\\
~~~~$i$\dotfill &Inclination (Degrees)\dotfill & $86.48^{+0.15}_{-0.20}$& $88.65^{+0.90}_{-0.92}$& $88.1^{+1.3}_{-1.2}$& $86.42^{+1.4}_{-0.85}$& $83.37^{+0.55}_{-0.75}$& $88.16^{+1.1}_{-0.80}$\\
~~~~$e$\dotfill &Eccentricity \dotfill & $0.052^{+0.062}_{-0.037}$& $0.394^{+0.035}_{-0.038}$& $0.208^{+0.034}_{-0.047}$& $0.057^{+0.068}_{-0.040}$& $0.117^{+0.10}_{-0.079}$& $0.195^{+0.043}_{-0.040}$\\
~~~~$\tau_{\rm circ}^\pi$\dotfill &Tidal circularization timescale (Gyr)\dotfill & $0.74^{+0.13}_{-0.15}$& $20.6^{+12}_{-7.7}$& $240^{+72}_{-59}$& $0.60^{+0.18}_{-0.17}$& $0.129^{+0.061}_{-0.050}$& $410^{+150}_{-130}$\\
~~~$\omega_*$\dotfill &Argument of Periastron (Degrees)\dotfill & $21^{+57}_{-33}$& $91.8^{+9.9}_{-9.6}$& $93^{+11}_{-13}$& $96^{+83}_{-89}$& $31^{+98}_{-36}$& $-20^{+18}_{-17}$\\
~~~~$T_{eq}$\dotfill &Equilibrium temperature (K)\dotfill & $962.2^{+8.0}_{-7.9}$& $1166^{+18}_{-16}$& $1484^{+16}_{-14}$& $1802^{+60}_{-54}$& $1580\pm27$& $1595^{+45}_{-42}$\\
~~~~$K$\dotfill &RV semi-amplitude (m/s)\dotfill & $146\pm11$& $341^{+28}_{-27}$& $350^{+23}_{-21}$& $291\pm36$& $105\pm21$& $300^{+23}_{-24}$\\
~~~~$R_P/R_*$\dotfill &Radius of planet in stellar radii \dotfill & $0.13272^{+0.00072}_{-0.00071}$& $0.07469^{+0.00052}_{-0.00044}$& $0.03996^{+0.00044}_{-0.00040}$& $0.0816\pm0.0012$& $0.10693^{+0.00095}_{-0.00090}$& $0.04333^{+0.00058}_{-0.00051}$\\
~~~~$a/R_*$\dotfill &Semi-major axis in stellar radii \dotfill& $12.28^{+0.27}_{-0.31}$& $13.12\pm0.48$& $8.65^{+0.22}_{-0.23}$& $6.76^{+0.26}_{-0.31}$& $7.91^{+0.33}_{-0.37}$& $10.63^{+0.58}_{-0.55}$\\
~~~~$Depth$\dotfill &Flux decrement at mid transit \dotfill & $0.01761\pm0.00019$& $0.005578^{+0.000078}_{-0.000065}$& $0.001597^{+0.000035}_{-0.000032}$& $0.00666^{+0.00020}_{-0.00019}$& $0.01143^{+0.00020}_{-0.00019}$& $0.001877^{+0.000050}_{-0.000044}$\\
~~~~$Depth_{\rm TESS}$\dotfill &Flux decrement at mid transit for \tess\ \dotfill & $0.01802\pm0.00014$& $0.006324^{+0.000080}_{-0.000079}$& ---& $0.00728^{+0.00018}_{-0.00017}$& $0.01046^{+0.00015}_{-0.00014}$& $0.002013\pm0.000044$\\
~~~~$\tau$\dotfill &Ingress/egress transit duration (days)\dotfill & $0.01918^{+0.00065}_{-0.00064}$& $0.01078^{+0.00092}_{-0.00036}$& $0.01260^{+0.0016}_{-0.00063}$& $0.0140\pm0.0019$& $0.0345^{+0.0023}_{-0.0022}$& $0.0154^{+0.0030}_{-0.0017}$\\
~~~~$T_{14}$\dotfill &Total transit duration (days)\dotfill & $0.08128^{+0.00057}_{-0.00056}$& $0.15003^{+0.00099}_{-0.00084}$& $0.3103^{+0.0021}_{-0.0019}$& $0.1548^{+0.0017}_{-0.0016}$& $0.1027\pm0.0010$& $0.3266^{+0.0031}_{-0.0027}$\\
~~~~$b$\dotfill &Transit Impact parameter \dotfill & $0.7393^{+0.0083}_{-0.0088}$& $0.19^{+0.14}_{-0.13}$& $0.23\pm0.16$& $0.415^{+0.098}_{-0.18}$& $0.8636^{+0.0071}_{-0.0082}$& $0.35^{+0.16}_{-0.21}$\\
~~~~$T_{S,14}$\dotfill &Total eclipse duration (days)\dotfill & $0.08168^{+0.00060}_{-0.00077}$& $0.314^{+0.041}_{-0.058}$& $0.452^{+0.041}_{-0.057}$& $0.1606^{+0.023}_{-0.0089}$& $0.0965^{+0.0072}_{-0.024}$& $0.292^{+0.030}_{-0.028}$\\
~~~~$\rho_P$\dotfill &Density (cgs)\dotfill & $1.23^{+0.13}_{-0.12}$& $3.74^{+0.51}_{-0.46}$& $5.34^{+0.59}_{-0.54}$& $1.67^{+0.31}_{-0.29}$& $0.370^{+0.092}_{-0.081}$& $6.1^{+1.2}_{-1.1}$\\
~~~~$logg_P$\dotfill &Surface gravity \dotfill & $3.387^{+0.038}_{-0.041}$& $3.899^{+0.047}_{-0.050}$& $4.057^{+0.038}_{-0.039}$& $3.630^{+0.065}_{-0.072}$& $3.032^{+0.088}_{-0.10}$& $4.082^{+0.059}_{-0.063}$\\
~~~~$T_S$\dotfill &Time of eclipse (\bjdtdb)\dotfill & $2458898.100^{+0.17}_{-0.089}$& $2458694.65^{+0.40}_{-0.41}$& $2459018.34^{+0.28}_{-0.27}$& $2458794.241^{+0.084}_{-0.099}$& $2458817.94^{+0.37}_{-0.23}$& $2459200.96^{+0.23}_{-0.24}$\\
~~~~$e\cos{\omega_*}$\dotfill & \dotfill & $0.036^{+0.070}_{-0.038}$& $-0.012^{+0.065}_{-0.067}$& $-0.010^{+0.042}_{-0.040}$& $-0.002^{+0.039}_{-0.046}$& $0.051^{+0.15}_{-0.094}$& $0.176^{+0.034}_{-0.035}$\\
~~~~$e\sin{\omega_*}$\dotfill & \dotfill & $0.015^{+0.029}_{-0.021}$& $0.388^{+0.035}_{-0.038}$& $0.203^{+0.034}_{-0.047}$& $0.022^{+0.076}_{-0.039}$& $0.035^{+0.054}_{-0.042}$& $-0.064^{+0.059}_{-0.067}$\\
~~~~$d/R_*$\dotfill &Separation at mid transit \dotfill & $12.04^{+0.53}_{-0.68}$& $8.00^{+0.75}_{-0.71}$& $6.91^{+0.49}_{-0.43}$& $6.58^{+0.48}_{-0.68}$& $7.48^{+0.69}_{-0.77}$& $10.97^{+1.1}_{-0.97}$\\
\hline
\end{tabular}
\begin{flushleft}
  \footnotesize{
    \textbf{\textsc{NOTES:}}\\
See Table 3 in \citet{Eastman:2019} for a detailed description of all derived and fitted parameters.\\
$^{**}$T$_C$ prior comes from analysis of the WASP photometry (see \S\ref{sec:wasp}). We note that this time is in HJD$_{\rm TDB}$ while all data files and results here are $\bjdtdb$. The difference between these two time systems is on the order of seconds while the precision on T$_C$ used as a prior is on order of minutes, and therefore has no influence on the results. \\
$^{\prime}$We assume the TESS correction for blending is much better than 10\%. We use a prior of 10\% of the determined blending from TICv8 \citep{Stassun:2018_TIC}.\\
$^\dagger$The initial metallicity is the metallicity of the star when it was formed.\\
$^\ddagger$The Equal Evolutionary Point corresponds to static points in a stars evolutionary history when using the MIST isochrones and can be a proxy for age. See \S2 in \citet{Dotter:2016} for a more detailed description of EEP.\\
$^\star$Optimal time of conjunction minimizes the covariance between $T_C$ and Period. This is the transit mid-point. \\
$^\pi$The tidal quality factor (Q$_S$) is assumed to be 10$^6$.\\ }
 \end{flushleft}
\label{tab:exofast_stellar}
\end{table*}
\begin{table*}
\centering
\scriptsize
\caption{Median values and 68\% confidence intervals for the global models}
\begin{tabular}{llcccccc}
  \hline
  \hline
TOI-1811&&\\
\multicolumn{2}{l}{Wavelength Parameters:}&B&R&g'&i'\\
&&r'&z'&TESS&V\smallskip\\
~~~~$u_{1}$\dotfill &linear limb-darkening coeff \dotfill &$0.944\pm0.037$&$0.665^{+0.045}_{-0.046}$&$0.941^{+0.041}_{-0.042}$&$0.482^{+0.023}_{-0.022}$\\
&&$0.593\pm0.042$&$0.394\pm0.033$&$0.417\pm0.039$&$0.746\pm0.050$\\
~~~~$u_{2}$\dotfill &quadratic limb-darkening coeff \dotfill &$-0.108\pm0.037$&$0.203^{+0.046}_{-0.047}$&$-0.032^{+0.045}_{-0.046}$&$0.171\pm0.022$\\
&&$0.090^{+0.044}_{-0.045}$&$0.196\pm0.033$&$0.130\pm0.043$&$0.055^{+0.050}_{-0.051}$\\
~~~~$A_D$\dotfill &Dilution from neighboring stars \dotfill &--&--&--&--\\
&&--&--&$-0.00001\pm0.00028$&--\\
\smallskip\\\multicolumn{2}{l}{Telescope Parameters:}&TRES\smallskip\\
~~~~$\gamma_{\rm rel}$\dotfill &Relative RV Offset (m/s)\dotfill &$-187.0^{+9.1}_{-9.7}$\\
~~~~$\sigma_J$\dotfill &RV Jitter (m/s)\dotfill &$17^{+14}_{-17}$\\
~~~~$\sigma_J^2$\dotfill &RV Jitter Variance \dotfill &$300^{+690}_{-320}$\\
\smallskip\\\multicolumn{2}{l}{Transit Parameters:}&TESS &LCOHAL0m4 UT 2020-04-23 (i')&LCOSSO0m4 UT 2020-04-23 (i')&BYU UT 2020-04-27 (R)\\
&ULMT UT 2020-04-27 (i') & Montcabrer UT 2020-04-30 (i')&BYU12in UT 2020-05-08 (V)&ElSauce UT 2020-05-12 (B)&LCOSSO1m UT 2021-02-25 (B)\\
&LCOSSO1m UT 2021-02-25 (z')&MUSCAT2 UT 2021-06-05 (g')&MUSCAT2 UT 2021-06-05 (i')&MUSCAT2 UT 2021-06-05 (r')&MUSCAT2 UT 2021-06-05 (z')\smallskip\\
~~~~$\sigma^{2}$\dotfill &Added Variance$\times$10$^{-5}$ \dotfill &$0.034^{+0.19}_{-0.16}$&$5.8^{+1.2}_{-1.0}$&$1.11^{+0.60}_{-0.49}$&$0.082^{+0.067}_{-0.057}$\\
&$0.098^{+0.034}_{-0.027}$& $0.34^{+0.19}_{-0.16}$&$2.10^{+0.39}_{-0.33}$&$1.18^{+0.53}_{-0.42}$&$0.20^{+0.13}_{-0.099}$\\
&$1.18^{+0.31}_{-0.25}$&$-0.027^{+0.055}_{-0.049}$&$0.061^{+0.057}_{-0.051}$&$-0.028^{+0.046}_{-0.043}$&$-0.079^{+0.054}_{-0.058}$\\
\smallskip\smallskip~~~~$F_0$\dotfill &Baseline flux \dotfill &$1.000031\pm0.000060$&$1.00236^{+0.00081}_{-0.00079}$&$1.01023^{+0.00068}_{-0.00066}$&$1.00007\pm0.00021$\\
&$0.99881\pm0.00016$&$1.00856\pm0.00034$&$1.00033\pm0.00046$&$0.99772\pm0.00061$&$0.99526^{+0.00033}_{-0.00034}$\\
&$0.99939^{+0.00049}_{-0.00050}$&$1.00004\pm0.00019$&$1.00016\pm0.00017$&$0.99991\pm0.00015$&$1.00056\pm0.00019$\\
\smallskip\smallskip~~~~$C_{0}$\dotfill &Additive detrending coeff \dotfill &--&$0.0074\pm0.0023$&$-0.0036\pm0.0022$&$0.00381\pm0.00052$\\
&$0.00512\pm0.00041$&$0.00186^{+0.00061}_{-0.00063}$&$-0.0039^{+0.0012}_{-0.0013}$&$0.0053^{+0.0018}_{-0.0019}$&$0.00816^{+0.00087}_{-0.00088}$\\
&$0.0010\pm0.0013$&--&--&--&--\\
\hline
  \hline
TOI-2025&&\\
\multicolumn{2}{l}{Wavelength Parameters:}&B&Kepler&R&g'\\
&&i'&TESS&V\smallskip\\
~~~~$u_{1}$\dotfill &linear limb-darkening coeff \dotfill &$0.611\pm0.053$&$0.416\pm0.036$&$0.346\pm0.050$&$0.556^{+0.055}_{-0.054}$\\
&&$0.283\pm0.037$&$0.247\pm0.026$&$0.432\pm0.039$\\
~~~~$u_{2}$\dotfill &quadratic limb-darkening coeff \dotfill &$0.178\pm0.053$&$0.310\pm0.035$&$0.300^{+0.049}_{-0.050}$&$0.236^{+0.051}_{-0.052}$\\
&&$0.296\pm0.036$&$0.273\pm0.028$&$0.277\pm0.037$\\
~~~~$A_D$\dotfill &Dilution from neighboring stars \dotfill &--&--&--&--\\&&--&$0.00001^{+0.00026}_{-0.00025}$&--\\
\smallskip\\\multicolumn{2}{l}{Telescope Parameters:}&TRES\smallskip\\
~~~~$\gamma_{\rm rel}$\dotfill &Relative RV Offset (m/s)\dotfill &$189\pm19$\\
~~~~$\sigma_J$\dotfill &RV Jitter (m/s)\dotfill &$61^{+22}_{-16}$\\
~~~~$\sigma_J^2$\dotfill &RV Jitter Variance \dotfill &$3800^{+3200}_{-1800}$\\
\smallskip\\\multicolumn{2}{l}{Transit Parameters:}& TESS S14,18,19,20,24,25,26&LCOTFN UT 2020-06-26 (g')&LCOTFN UT 2020-06-26 (i')&SCT UT 2020-06-26 (TESS)\\
&TESS S40&KeplerCam UT 2021-05-12 (B)&KeplerCam UT 2021-05-12 (i')&GMU UT 2021-05-20 (V)&CRCAO UT 2021-05-21 (R)\\
&MORP UT 2021-09-30 (Kepler)&MORP UT 2021-10-18 (Kepler)&Conti UT 2021-12-19 (V)\smallskip\\
~~~~$\sigma^{2}$\dotfill &Added Variance$\times$10$^{-5}$ \dotfill &$-0.0064^{+0.0012}_{-0.0011}$&$1.07^{+0.31}_{-0.25}$&$1.05^{+0.30}_{-0.24}$&$2.65^{+0.24}_{-0.22}$\\
&$-0.0035^{+0.010}_{-0.0097}$&$0.527^{+0.066}_{-0.057}$&$1.34^{+0.16}_{-0.14}$&$1.33^{+0.16}_{-0.14}$&$0.397^{+0.070}_{-0.060}$\\
&$0.554^{+0.053}_{-0.049}$&$0.370^{+0.048}_{-0.044}$&$1.78^{+0.33}_{-0.31}$\\
~~~~$F_0$\dotfill &Baseline flux \dotfill &$1.000005\pm0.000021$&$0.99958^{+0.00046}_{-0.00045}$&$0.99954\pm0.00044$&$0.99977\pm0.00028$\\
&$1.000892\pm0.000042$&$1.00386\pm0.00019$&$1.00379\pm0.00029$&$1.00051^{+0.00039}_{-0.00038}$&$1.00387\pm0.00019$\\
&$1.00023\pm0.00014$&$0.99799\pm0.00012$&$1.00293\pm0.00029$\\
~~~~$C_{0}$\dotfill &Additive detrending coeff \dotfill &--&$0.00283\pm0.00096$&$0.00147\pm0.00094$&$-0.00070\pm0.00080$\\
&--&$-0.00278\pm0.00045$&$-0.00195\pm0.00068$&$-0.00026^{+0.00088}_{-0.00089}$&$0.00356^{+0.00041}_{-0.00042}$\\
&--&--&$-0.00577^{+0.00086}_{-0.00085}$\\
\hline
\hline
TOI-2145&&\\
\smallskip\\\multicolumn{2}{l}{Wavelength Parameters:}&i'&r'&TESS\smallskip\\
~~~~$u_{1}$\dotfill &linear limb-darkening coeff \dotfill &$0.242\pm0.051$&$0.316^{+0.051}_{-0.050}$&$0.220\pm0.029$\\
~~~~$u_{2}$\dotfill &quadratic limb-darkening coeff \dotfill &$0.314\pm0.050$&$0.310\pm0.050$&$0.297^{+0.034}_{-0.033}$\\
\smallskip\\\multicolumn{2}{l}{Telescope Parameters:}&MINERVAT1&MINERVAT2&MINERVAT3&TRES\smallskip\\
~~~~$\gamma_{\rm rel}$\dotfill &Relative RV Offset (m/s)\dotfill &$61\pm41$&$-13^{+27}_{-25}$&$-50^{+140}_{-180}$&$-194\pm16$\\
~~~~$\sigma_J$\dotfill &RV Jitter (m/s)\dotfill &$167^{+43}_{-35}$&$71^{+32}_{-27}$&$220^{+410}_{-220}$&$35\pm22$\\
~~~~$\sigma_J^2$\dotfill &RV Jitter Variance \dotfill &$28000^{+16000}_{-11000}$&$5100^{+5600}_{-3100}$&$47000^{+340000}_{-55000}$&$1300^{+2100}_{-1100}$\\
\smallskip\\\multicolumn{2}{l}{Transit Parameters:}&TESS S26/27&TESS S40&ULMoore UT 2021-09-07 (i')&CRCAO UT 2021-09-07 (r')\smallskip\\
~~~~$\sigma^{2}$\dotfill &Added Variance$\times$10$^{-5}$ \dotfill &$0.00435^{+0.00087}_{-0.0083}$&$0.00273^{+0.00075}_{-0.00073}$&$1.74^{+0.23}_{-0.20}$&$1.221^{+0.092}_{-0.084}$\\
~~~~$F_0$\dotfill &Baseline flux \dotfill &$1.000900\pm0.000012$&$1.000627\pm0.000011$&$1.00087\pm0.00036$&$1.00066\pm0.00017$\\
~~~~$C_{0}$\dotfill &Additive detrending coeff \dotfill &--&--&$-0.00136^{+0.00070}_{-0.00071}$&$-0.00398\pm0.00040$\\
\hline
\hline
\label{tab:exofast_other1}
\end{tabular}
\begin{flushleft}
  \end{flushleft}
\end{table*}

\begin{table*}
\centering
\scriptsize
\caption{Median values and 68\% confidence intervals for the global models}
\begin{tabular}{llcccccc}
  \hline
  \hline
  TOI-2152&&\\
\multicolumn{2}{l}{Wavelength Parameters:}&B&R&g'&i'\\
&&TESS\smallskip\\
~~~~$u_{1}$\dotfill &linear limb-darkening coeff \dotfill &$0.484^{+0.067}_{-0.061}$&$0.262^{+0.047}_{-0.044}$&$0.423^{+0.067}_{-0.064}$&$0.207^{+0.057}_{-0.056}$\\
&&$0.205^{+0.039}_{-0.040}$\\
~~~~$u_{2}$\dotfill &quadratic limb-darkening coeff \dotfill &$0.265^{+0.051}_{-0.054}$&$0.337\pm0.037$&$0.287^{+0.056}_{-0.057}$&$0.327\pm0.050$\\
&&$0.330\pm0.049$\\
~~~~$A_D$\dotfill &Dilution from neighboring stars \dotfill &--&--&--&--\\
&&$-0.001\pm0.018$\\
\smallskip\\\multicolumn{2}{l}{Telescope Parameters:}&TRES\smallskip\\
~~~~$\gamma_{\rm rel}$\dotfill &Relative RV Offset (m/s)\dotfill &$207^{+29}_{-28}$\\
~~~~$\sigma_J$\dotfill &RV Jitter (m/s)\dotfill &$83^{+37}_{-26}$\\
~~~~$\sigma_J^2$\dotfill &RV Jitter Variance \dotfill &$7000^{+7500}_{-3700}$\\
\smallskip\\\multicolumn{2}{l}{Transit Parameters:}&TESS &OWL UT 2020-08-17 (R)&WaffelowCreek UT 2020-10-11 (g')&WaffelowCreek UT 2020-10-11 (i')\\
&CALOU UT 2020-11-24 (B)&Kourovka UT 2020-12-10 (B)&CRCAO UT 2021-06-28 (R)\smallskip\\
~~~~$\sigma^{2}$\dotfill &Added Variance$\times$10$^{-5}$ \dotfill &$-0.00131^{+0.00098}_{-0.00092}$&$1.16^{+0.21}_{-0.19}$&$0.56^{+0.22}_{-0.18}$&$0.68^{+0.18}_{-0.15}$\\
&$0.73^{+0.20}_{-0.17}$&$0.67^{+0.15}_{-0.12}$&$1.23^{+0.15}_{-0.13}$\\
~~~~$F_0$\dotfill &Baseline flux \dotfill &$1.000023\pm0.000017$&$1.00036\pm0.00030$&$1.00344\pm0.00038$&$1.00342\pm0.00035$\\
&$0.99928^{+0.00035}_{-0.00034}$&$1.00137\pm0.00032$&$1.00032\pm0.00028$\\
~~~~$C_{0}$\dotfill &Additive detrending coeff \dotfill &--&$0.00283\pm0.00066$&$-0.00219^{+0.00094}_{-0.00093}$&$-0.00104^{+0.00084}_{-0.00085}$\\
&$0.00109\pm0.00070$&$-0.00132^{+0.00052}_{-0.00051}$&$0.00008\pm0.00044$\\
\hline
\hline
TOI-2154&&\\
\multicolumn{2}{l}{Wavelength Parameters:}&B&I&Kepler&z'&TESS\smallskip\\
~~~~$u_{1}$\dotfill &linear limb-darkening coeff \dotfill &$0.493\pm0.056$&$0.211\pm0.052$&$0.325\pm0.052$&$0.195\pm0.038$&$0.264\pm0.049$\\
~~~~$u_{2}$\dotfill &quadratic limb-darkening coeff \dotfill &$0.205\pm0.054$&$0.298\pm0.050$&$0.306^{+0.049}_{-0.050}$&$0.304\pm0.035$&$0.329\pm0.048$\\
~~~~$A_D$\dotfill &Dilution from neighboring stars \dotfill &--&--&--&--&$0.00001\pm0.00055$\\
\smallskip\\\multicolumn{2}{l}{Telescope Parameters:}&TRES\smallskip\\
~~~~$\gamma_{\rm rel}$\dotfill &Relative RV Offset (m/s)\dotfill &$10^{+20}_{-18}$\\
~~~~$\sigma_J$\dotfill &RV Jitter (m/s)\dotfill &$32^{+26}_{-32}$\\
~~~~$\sigma_J^2$\dotfill &RV Jitter Variance \dotfill &$1000^{+2300}_{-1100}$\\
\smallskip\\\multicolumn{2}{l}{Transit Parameters:}& TESS&V390m4 UT 2020-08-18 (I)&OPM UT 2020-10-29 (z')&CALOU UT 2020-11-23 (B)\\
&&LCO McD UT 2020-12-03 (z')&MSU UT 2021-10-24 (Kepler)\smallskip\\
~~~~$\sigma^{2}$\dotfill &Added Variance$\times$10$^{-5}$ \dotfill &$-0.1708^{+0.0014}_{-0.0013}$&$2.42^{+0.59}_{-0.48}$&$1.6^{+1.8}_{-0.000014}$&$0.46^{+0.13}_{-0.11}$\\
&&$0.78^{+0.16}_{-0.13}$&$0.139^{+0.044}_{-0.036}$\\
~~~~$F_0$\dotfill &Baseline flux \dotfill &$1.000067\pm0.000021$&$0.99986\pm0.00064$&$0.9959\pm0.0012$&$0.99915^{+0.00027}_{-0.00028}$\\
&&$1.00037\pm0.00033$&$0.99992^{+0.00018}_{-0.00017}$\\
~~~~$C_{0}$\dotfill &Additive detrending coeff \dotfill &--&$0.0004\pm0.0011$&$0.0010\pm0.0022$&--\\
&&$0.00116\pm0.00078$&$0.00064\pm0.00041$\\
~~~~$M_{0}$\dotfill &Multiplicative detrending coeff \dotfill &--&--&--&$0.00160\pm0.00054$\\
&&--&--\\
\hline
\hline
TOI-2497&&\\
\multicolumn{2}{l}{Wavelength Parameters:}&TESS\smallskip\\
~~~~$u_{1}$\dotfill &linear limb-darkening coeff \dotfill &$0.156^{+0.035}_{-0.034}$\\
~~~~$u_{2}$\dotfill &quadratic limb-darkening coeff \dotfill &$0.327\pm0.036$\\
\smallskip\\\multicolumn{2}{l}{Telescope Parameters:}&CHIRON&MINERVAT3&MINERVAT4&MINERVAT5\\
&&MINERVAT6&TRES\smallskip\\
~~~~$\gamma_{\rm rel}$\dotfill &Relative RV Offset (m/s)\dotfill &$43\pm24$&$55864^{+83}_{-79}$&$56050^{+110}_{-100}$&$56280\pm100$\\
&$56154^{+73}_{-80}$&$-342^{+27}_{-26}$\\
~~~~$\sigma_J$\dotfill &RV Jitter (m/s)\dotfill &$78^{+30}_{-27}$&$210^{+83}_{-69}$&$251^{+120}_{-99}$&$80^{+210}_{-81}$\\
&$0.00^{+250}_{-0.00}$&$83^{+32}_{-28}$\\
~~~~$\sigma_J^2$\dotfill &RV Jitter Variance \dotfill &$6100^{+5500}_{-3500}$&$44000^{+42000}_{-24000}$&$63000^{+75000}_{-40000}$&$7000^{+76000}_{-35000}$\\
&$0^{+61000}_{-29000}$&$6900^{+6300}_{-3800}$\\
\smallskip\\\multicolumn{2}{l}{Transit Parameters:}&TESS S6 &TESS S33\smallskip\\
~~~~$\sigma^{2}$\dotfill &Added Variance$\times$10$^{-5}$ \dotfill &$0.015^{+0.00013}_{-0.00010}$&$0.0018\pm0.016$\\
~~~~$F_0$\dotfill &Baseline flux \dotfill &$1.000031^{+0.000032}_{-0.000033}$&$1.000258\pm0.000019$\\
\hline
\hline
\label{tab:exofast_other2}
\end{tabular}
\begin{flushleft}
  \end{flushleft}
\end{table*}

\section{Discussion}
\label{sec:discussion}
The combination of precision, baseline, and cadence of \tess\ will provide the ability to create a magnitude-complete, self-consistent catalog of exoplanetary systems to investigate questions about formation and evolution, and directly test tentative trends seen in the current population \citep{Nelson:2017, Rodriguez:2021, IkwutUkwa:2022}. These six new hot and warm giant planets increase the current sample of systems with precise mass and eccentricity measurements. We first review our results on each system and then discuss the impact \tess\ has made on the field of giant exoplanets. In all six systems, we see no significant inflation ($R_P$ $>$1.5 \rsun). We also see some significant reddening for TOI-2152 and TOI-2497 from our global fit (see Table \ref{tab:exofast_stellar}).

\subsection{Review of Six New Discoveries}
Orbiting an early K-star, TOI-1811 b is a hot Jupiter on a 3.71 day orbital period that shows no signs of inflation relative to the known population (R$_P$ = $0.994^{+0.025}_{-0.023}$\rj\ and M$_P$ = $0.972^{+0.076}_{-0.078}$\mj). The host star has a relatively high metallicity (\feh\ = $0.306^{+0.076}_{-0.077}$ dex), and the lack of a significant eccentricity is consistent with the very short tidal circularization timescale of $740^{+13}_{-15}$ Myr \citep{Adams:2006} and that the host star parameters suggest a main-sequence star with an age well above this. 

TOI-2025 b is a super Jupiter mass (M$_P$ = $3.60\pm0.33$\mj) planet on an 8.872 day orbital period around an early-G star. We detect a moderate, but significant eccentricity, e = $0.394^{+0.035}_{-0.038}$. Given the long circularization timescale (see Table \ref{tab:exofast_stellar}) and the detected eccentricity, it is possible that TOI-2025 b migrated to its current location through dynamical interactions \citep[e.g.,][]{Dawson:2010}. 

Orbiting a bright ($G$ = 8.94$\pm$0.02 mag), sub-giant ($\log{g}$ = $3.794^{+0.023}_{-0.027}$ cgs), TOI-2145 is a massive (M$_P$ = $5.26^{+0.38}_{-0.37}$\mj) warm Jupiter on an eccentric (e = $0.208^{+0.034}_{-0.047}$) on a 10.261 day orbit. Of the known transiting planets to date, TOI-2145 b joins only five other known planets to have a mass above 3\mj\ and orbit a subgiant ($\log{g}$ $<$ 4.0 cgs), but it orbits the brightest star of that group, a valuable aspect for future detailed characterization. 

TOI-2152A b and TOI-2154 b are both hot Jupiters orbiting similar main-sequence F-stars at similar distances from the Sun. TOI-2152A b is a massive Jupiter (M$_P$ = $2.83^{+0.38}_{-0.37}$\mj) while TOI-2154 b is only  $0.92^{+0.19}_{-0.18}$\mj. We see no evidence of any significant eccentricity (TOI-2152A b $e$ = $0.057^{+0.068}_{-0.040}$, TOI-2154 b  $e$ = $0.117^{+0.10}_{-0.079}$) from our results but note that these two planets provide a nice comparative study since their host stars and the planets share many similar characteristics, but exhibit a significant difference in the planet's mass.

The last system in our sample is TOI-2497 b, another very massive (M$_P$ = $5.21\pm0.52$\mj) warm Jupiter on a 10.656 day orbital period. Its host star, TOI-2497, is a rapidly rotating ($\vsini$ = 39.6$\pm$1.0 \kms) early F-star ($T_{\rm eff}$ = $7360^{+320}_{-300}$, that has possibly left the main sequence ($\log{g}$ =  $3.962^{+0.050}_{-0.049}$ cgs). The host star is also bright ($G$ = 9.47$\pm$0.02 mag), and combined with the rapid rotation, TOI-2497 b is an excellent target for future Doppler spectroscopy, using observations of the Rossiter McLauglin effect \citep{Rossiter:1924, McLaughlin:1924} or Doppler tomography \citep[e.g.,][]{Miller:2010, Johnson:2014, Zhou:2016} to measure the projected spin-orbit alignment of the planet's orbit. 

\subsection{\tess's impact on Giant Planets}
\begin{figure*}
	\centering\vspace{.0in}
	\includegraphics[width=0.99\linewidth, trim={0 0 0 0}, clip]{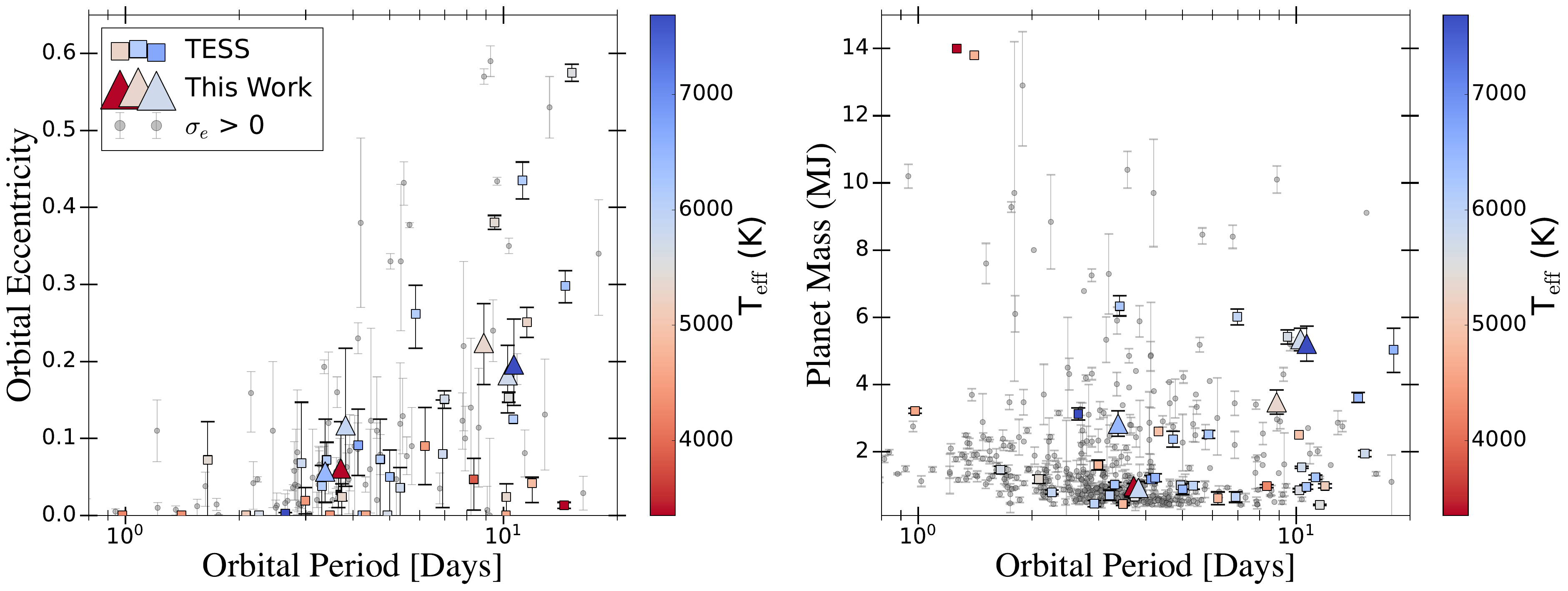}
\caption{(Left) The eccentricity and log of the orbital period of all known giant planets with a mass greater than 0.4\mj with period between 0.8 and 30 days. The systems with a measured eccentricity from the NASA Exoplanet Archive (NEA) are shown as grey circles with errors.  (Right) The mass and log of the orbital period of all known transiting giant planets. In both figures the \tess-discovered systems (including the ones presented in this work) are the squares colored by the host star's effective temperature (with those from this work are displayed with a triangle symbol), showing the diversity of the host stars in the \tess\ giant planet sample. } 
	\label{fig:discussion} 
\end{figure*}

As NASA's \tess\ mission continues to observe, it is expected to discover thousands of giant planets over its lifetime \citep{Sullivan:2015, Barclay:2018}, while providing great value to already known systems \citep{IkwutUkwa:2020, Edwards:2021, Kane:2021}. This is highly dependent on the number of extended missions that \tess\ is given. Even in the $\sim$4 years since its launch, \tess\ has discovered over 200 planets\footnote{\url{https://exoplanetarchive.ipac.caltech.edu}, accessed April 2022}, of which 47 are above 0.4\mj, nearly 10\% of the known transiting giant planet population (See Figure \ref{fig:discussion}). As multiple efforts, including ours, continue to confirm and characterize new transiting giant planets, it will lead to a magnitude-complete, self-consistent sample of planet properties \citep{Zhou:2019, Yee:2021}. 

There is an obvious trend in the eccentricity distribution of giant planets, where long period giant planets tend to have a wider distribution of orbital eccentricities than shorter period systems, possibly indicative of the system's migration history. If a planet migrates to a close-in configuration through dynamical interactions with other bodies, it can result in a highly eccentric and/or misaligned orbit \citep{Rasio:1996,Wu:2011}. Specifically, looking at Figure \ref{fig:discussion}, we see that the eccentricity range appears to broaden beyond an orbital period of $\sim$3 days. We note that many components of a planet's formation and evolutionary history are incorporated into this distribution, and a proper analysis of the population as a function of host star parameters is warranted prior to drawing any conclusions. This trend is also seen for brown dwarfs, indicating that more massive systems may undergo migration scenarios similar to planets \citep{Carmichael:2021}. 

Another possible piece of the puzzle is that a tentative trend has emerged where longer period hot Jupiters ($>$5 days) are more massive than shorter period ones \citep{IkwutUkwa:2022}. Unfortunately, the lack of homogenity of the current exoplanet population makes any observed trends difficult to interpret since they may only manifest due to the different assumptions and analysis techniques used. More importantly, Figure \ref{fig:discussion} shows the large impact \tess\ is making on the field of giant planets purely from the large number of Jovian-sized planets it has discovered to date, with many of them on longer orbital periods (P $>$ 5 days) where the ground-based transit surveys struggled due to poor duty cycles \citep{Gaudi:2005}. With the expectation of hundreds of additional discoveries as \tess\ continues to scan the entire sky, the community will have a large number of systems to consider for future detailed characterization using ongoing and future facilities like the James Webb Space Telescope (JWST), the Atmospheric Remote-sensing Infrared Exoplanet Large-survey (ARIEL, \citealp{Tinetti:2016}), and future 30-meter class ground-based telescopes. Future work should consider obtaining Doppler spectroscopy on TOI-2497 b to determine the orbital obliquity of the planet, a key aspect related to a planet's migration history.

\section{Conclusion}
\label{sec:conclusion}
Using a combination of photometric and spectroscopic observations, we present the discovery of six new hot and warm giant planets (TOI-1811 b, TOI-2025 b, TOI-2145 b, TOI-2152A b, TOI-2154 b, and TOI-2497 b). These systems increase the number of giant planets discovered by \tess\ to date and are a part of a larger effort to create a complete sample of systems brighter than $G$ $<$ 12.5 in support of future population studies. Of the six systems presented here, we note a few interesting aspects. First, TOI-2145 is a bright ($G$ = 8.94$\pm$0.02 mag), sub-giant ($\log{g}$ = $3.798^{+0.023}_{-0.026}$ cgs) with a 10.26 day period and a $\sim$5\mj\ planet. Interestingly, we see no signs of inflation from the measured radius of TOI-2145 b, but it is important to note that hot Jupiters discovered around evolved stars suggest planets may re-inflate in the post-main sequence phase \citep{Almenara:2015, Grunblatt:2016, Hartman:2016, Stevens:2017, Komacek:2020}, when a warm Jupiter (like TOI-2145 b) will receive a similar amount of irradiation to that of a hot Jupiter \citep{Lopez:2016}. TOI-2152A b and TOI-2154 b are similar orbital period hot Jupiters that orbit similar hosts but the planets are $2.83^{+0.38}_{-0.37}$\mj\ and $0.92^{+0.19}_{-0.18}$\mj\ providing a nice opportunity for future comparative studies. TOI-2497 b orbits a massive, early F-star ($T_{\rm eff}$ = $7360^{+290}_{-270}$), and the combination of its host star's brightness ($G$ = 9.47$\pm$0.02 mag) and rotation period ($\vsini$ = 39.6$\pm$1.0 \kms) make it well-suited for orbital obliquity measurements through transit spectroscopy followup. \TESS\ continues to discover a wealth of transiting giant planets that may provide insight into their formation and evolutionary mechanisms.

\section*{Acknowledgements}

L.C, K.S., E.A., J.R., J.E.R., J.A.R., P.W., and E.Z. are grateful for support from NSF grants AST-1751874 and AST-1907790, along with a Cottrell Fellowship from the Research Corporation. CZ is supported by a Dunlap Fellowship at the Dunlap Institute for Astronomy \& Astrophysics, funded through an endowment established by the Dunlap family and the University of Toronto. T.H. acknowledges support from the European Research Council under the Horizon 2020 Framework Program via the ERC Advanced Grant Origins 83 24 28. J.V.S. acknowledges funding from the European Research Council (ERC) under the European Union’s Horizon 2020 research and innovation programme (project Four Aces; grant agreement No. 724427). P. R. acknowledges support from NSF grant No. 1952545. R.B.\ and A.J. acknowledges support from FONDECYT Project 11200751 and from CORFO project N$^\circ$14ENI2-26865. A.J.\, R.B.\, and M.H.\ acknowledge support from project IC120009 ``Millennium Institute of Astrophysics (MAS)'' of the Millenium Science Initiative, Chilean Ministry of Economy. The Pennsylvania State Uni- versity Eberly College of Science. The Center for Exoplanets and Habitable Worlds is supported by the Pennsylvania State Univer- sity, the Eberly College of Science, and the Pennsylvania Space Grant Consortium. K.K.M. gratefully acknowledges support from the NewYork CommunityTrust's Fund for Astrophysical Research. L.G. and A.G. are supported by NASA Massachusetts Space Grant Fellowships. E.W.G., M.E., and P.C. acknowledge support by Deutsche Forschungsgemeinschaft (DFG) grant  HA 3279/12-1  within the DFG Schwerpunkt SPP1992, Exploring the Diversity of Extrasolar Planets. B.S.G. was partially supported by the Thomas Jefferson Chair for Space Exploration at the Ohio State University. C.D. acknowledges support from the Hellman Fellows Fund and NASA XRP via grant 80NSSC20K0250. B.S.S., M.V.G. and A.A.B. acknowledge the support of Ministry of Science and Higher Education of the Russian Federation under the grant 075-15-2020-780 (N13.1902.21.0039). B.A. is supported by Australian Research Council Discovery Grant DP180100972.  T.R.B acknowledges support from the Australian Research Council (DP210103119). T.R.B acknowledges support from the Australian Research Council (DP210103119 and FL220100117)

We thank the CHIRON team members, including Todd Henry, Leonardo Paredes, Hodari James, Azmain Nisak,  Rodrigo Hinojosa, Roberto Aviles, Wei-Chun Jao and CTIO staffs, for their work in acquiring RVs with CHIRON at CTIO. This research has made use of SAO/NASA's Astrophysics Data System Bibliographic Services. This research has made use of the SIMBAD database, operated at CDS, Strasbourg, France. This work has made use of data from the European Space Agency (ESA) mission {\it Gaia} (\url{https://www.cosmos.esa.int/gaia}), processed by the {\it Gaia} Data Processing and Analysis Consortium (DPAC, \url{https://www.cosmos.esa.int/web/gaia/dpac/consortium}). Funding for the DPAC has been provided by national institutions, in particular the institutions participating in the {\it Gaia} Multilateral Agreement. This work makes use of observations from the LCO network. Based in part on observations obtained at the Southern Astrophysical Research (SOAR) telescope, which is a joint project of the Minist\'{e}rio da Ci\^{e}ncia, Tecnologia e Inova\c{c}\~{o}es (MCTI/LNA) do Brasil, the US National Science Foundation’s NOIRLab, the University of North Carolina at Chapel Hill (UNC), and Michigan State University (MSU).

Funding for the {\it TESS} mission is provided by NASA's Science Mission directorate. We acknowledge the use of public {\it TESS} Alert data from pipelines at the {\it TESS} Science Office and at the {\it TESS} Science Processing Operations Center. This research has made use of the NASA Exoplanet Archive and the Exoplanet Follow-up Observation Program website, which are operated by the California Institute of Technology, under contract with the National Aeronautics and Space Administration under the Exoplanet Exploration Program. This paper includes data collected by the {\it TESS} mission, which are publicly available from the Mikulski Archive for Space Telescopes (MAST). This paper includes observations obtained under Gemini program GN-2018B-LP-101. Resources supporting this work were provided by the NASA High-End Computing (HEC) Program through the NASA Advanced Supercomputing (NAS) Division at Ames Research Center for the production of the SPOC data products. This publication makes use of The Data \& Analysis Center for Exoplanets (DACE), which is a facility based at the University of Geneva (CH) dedicated to extrasolar planets data visualisation, exchange and analysis. DACE is a platform of the Swiss National Centre of Competence in Research (NCCR) PlanetS, federating the Swiss expertise in Exoplanet research. The DACE platform is available at \url{https://dace.unige.ch}.

Some of the data presented herein were obtained at the W. M. Keck Observatory, which is operated as a scientific partnership among the California Institute of Technology, the University of California and the National Aeronautics and Space Administration. The Observatory was made possible by the generous financial support of the W. M. Keck Foundation. The authors wish to recognize and acknowledge the very significant cultural role and reverence that the summit of Mauna Kea has always had within the indigenous Hawaiian community.  We are most fortunate to have the opportunity to conduct observations from this mountain.

\textsc{Minerva}-Australis is supported by Australian Research Council LIEF Grant LE160100001 (Discovery Grant DP180100972 and DP220100365) Mount Cuba Astronomical Foundation, and institutional partners University of Southern Queensland, UNSW Sydney, MIT, Nanjing University, George Mason University, University of Louisville, University of California Riverside, University of Florida, and The University of Texas at Austin. We respectfully acknowledge the traditional custodians of all lands throughout Australia, and recognise their continued cultural and spiritual connection to the land, waterways, cosmos, and community. We pay our deepest respects to all Elders, ancestors and descendants of the Giabal, Jarowair, and Kambuwal nations, upon whose lands the \textsc{Minerva}-Australis facility at Mt Kent is situated. 

Data presented herein were obtained at the MINERVA-Australis from telescope time allocated under the NN-EXPLORE program with support from the National Aeronautics and Space Administration.

MINERVA-North is a collaboration among the Harvard-Smithsonian Center for Astrophysics, The Pennsylvania State University, the University of Montana, the University of Southern Queensland, University of Pennsylvania, and George Mason University. It is made possible by generous contributions from its collaborating institutions and Mt. Cuba Astronomical Foundation, The David \& Lucile Packard Foundation, National Aeronautics and Space Administration (EPSCOR grant NNX13AM97A, XRP 80NSSC22K0233), the Australian Research Council (LIEF grant LE140100050), and the National Science Foundation (grants 1516242, 1608203, and 2007811).  

This article is based on observations made with the MuSCAT2 instrument, developed by ABC, at Telescopio Carlos Sánchez operated on the island of Tenerife by the IAC in the Spanish Observatorio del Teide. This work is partly financed by the Spanish Ministry of Economics and Competitiveness through grants PGC2018-098153-B-C31.The work of VK was supported by the Ministry of science and higher education of the Russian Federation, topic FEUZ-2020-0038.

This work is partly supported by JSPS KAKENHI Grant Number JP18H05439, JST CREST Grant Number JPMJCR1761. This article is based on observations made with the MuSCAT2 instrument, developed by ABC, at Telescopio Carlos Sánchez operated on the island of Tenerife by the IAC in the Spanish Observatorio del Teide. 

The Center for Exoplanets and Habitable Worlds and the Penn State Extraterrestrial Intelligence Center are supported by Penn State and the Eberly College of Science.

This paper was partially based on observations obtained at the OWL-Net system, which is operated by the Korea Astronomy and Space Science Institute (KASI).

\section*{Data Availability}
The \tess\ observations used in this paper (see \S\ref{sec:TESS}) and are shown in in Figure \ref{fig:fullLCs} are publicly available on the MAST\footnote{\url{https://mast.stsci.edu/}} archive. The photometric transit follow up observations from the SG1 working groups in TFOP (underlying data for Figure \ref{fig:transits1} and \ref{fig:transits2}) are publicly available on Exofop\footnote{\url{https://exofop.ipac.caltech.edu/tess/}}, along with the the AO and SPECKLE contrast curves and images discussed in \S2.6. The RV data (sample shown in Table \ref{tab:rv}) underlying this article (shown in Figure \ref{fig:RVs}) are available in the article and in its online supplementary material.
\\

Software Used: \texttt{EXOFASTv2} \citep{Eastman:2013, Eastman:2019}, AstroImageJ \citep{Collins:2017}, TAPIR \citep{Jensen:2013}, QLP Pipeline \citep{Huang:QLP}

Facilities: \tess, FLWO 1.5m (Tillinghast Reflector Echelle Spectrograph), 4.1-m Southern Astrophysical Research (SOAR), LCO 0.4m, LCO 1.0m, 2.2m telescope La Silla (Fiber-fed Extended Range Optical Spectrograph),  KECK (NIRC2), PALOMAR (PHARO), KELT, WASP, CTIO 1.5m (CHIRON), \textsc{MINERVA}-North,\textsc{Minerva}-Australis, GEMINI (NIRI), CMO 2.5m (SPP)

\bibliographystyle{mnras}

\bibliography{refs}

\section*{Affiliations}
\noindent
$^1$\msu\\
$^2$\cfa\\
$^3$\MIT\\
$^4$\usq\\
$^5$\nexsci\\
$^6$\Kutztown\\
$^7$\txamGP\\
$^{8}$\MunnerlynLab\\
$^{9}$\CGWA\\
$^{10}$\ucsc\\
$^{11}$\osu\\
$^{12}$\keele\\
$^{13}$\eso\\
$^{14}$\Pontificia\\
$^{15}$\Millennium\\
$^{16}$\DOF\\
$^{17}$\Kotizarovci\\
$^{18}$\louisville\\
$^{19}$\calou\\
$^{20}$\Villa\\\
$^{21}$\ElSauce\\
$^{22}$\byu\\
$^{23}$\wellesley\\
$^{24}$\Ural\\
$^{25}$\Oikaimeden\\
$^{26}$\liegeastrobio\\
$^{27}$\Sternberg\\
$^{28}$\psu\\
$^{29}$\psust\\
$^{30}$\aavso\\
$^{31}$\berkely\\
$^{32}$\SFASU\\
$^{33}$\albany\\
$^{34}$\citizen\\
$^{35}$\iac\\
$^{36}$\lalaguna\\
$^{37}$\waffelow\\
$^{38}$\MITEPS\\
$^{39}$\MITAA\\
$^{40}$\princeton\\
$^{41}$\ames\\
$^{42}$\swinburne
$^{43}$\gmu\\
$^{44}$\warwick\\
$^{45}$\wisconsin\\
$^{46}$\sifa\\
$^{47}$\UPenn\\
$^{48}$\utaustin\\
$^{49}$\montana\\
$^{50}$\ctio\\
$^{51}$\utah\\
$^{52}$\Cadi\\
$^{53}$\vanderbilt\\
$^{54}$\Komaba\\
$^{55}$\Tsinghua\\
$^{56}$\maxplank\\
$^{57}$\ASTRAVEO\\
$^{58}$\riverside\\
$^{59}$\KASSI\\
$^{60}$\TJHS\\
$^{61}$\saao\\
$^{62}$\salt\\
$^{63}$\unc\\
$^{64}$\maurylewin\\
$^{65}$\nanjing\\
$^{66}$\astrobiojapan\\
$^{67}$\bern\\
$^{68}$\lehigh\\
$^{69}$\ucscchile\\
$^{70}$\umd\\
$^{71}$\Patashnick\\
$^{72}$\gemini\\
$^{73}$\fisk\\
$^{74}$\Uminn
$^{75}$\CAUP\\
$^{76}$\Faculdade\\
$^{77}$\unsw\\
$^{78}$\warwickceh\\
$^{79}$\PSUET\\
$^{80}$\Shanghai\\
%



\label{lastpage}
\end{document}